\documentclass[aps,10pt,prx,twocolumn,longbibliography]{revtex4-1}
\usepackage[latin1]{inputenc} 
\usepackage[T1]{fontenc}
\usepackage{lmodern}
\usepackage{graphicx}
\usepackage{amsmath,amssymb,color,bm}
\usepackage[normalem]{ulem}
\renewcommand{\d}{\mathrm{d}}

\begin{document}
\title{Ionic Coulomb blockade as a fractional Wien effect}
\author{Nikita Kavokine, Sophie Marbach, Alessandro Siria and Lyd\'eric Bocquet}\email{lyderic.bocquet@ens.fr}
\affiliation{Laboratoire de Physique de l'\'Ecole Normale Sup\'erieure, ENS, Universit\'e PSL, CNRS, Sorbonne Universit\'e, Universit\'e Paris-Diderot, Sorbonne Paris Cit\'e, Paris, France}

\begin{abstract}

Recent advances in nanofluidics have allowed exploration of ion transport down to molecular scale confinement, yet artificial porins are still far 
from reaching the advanced functionalities of biological ion machinery. Achieving single ion transport that is tuneable by an external gate -- the ionic analogue of electronic Coulomb blockade (CB) -- would open new avenues in this quest. However, an understanding of ionic CB beyond the electronic analogy is still lacking. 
Here we show that the many-body dynamics of ions in a charged nanochannel result in a quantised and strongly non-linear ionic transport, in full agreement with molecular simulations.
We find that ionic CB occurs when, upon sufficient confinement, oppositely charged ions form 'Bjerrum pairs', and the conduction proceeds through a mechanism reminiscent of Onsager's Wien effect. 
Our findings open the way to novel nanofluidic functionalities, such as an ionic-CB-based ion pump inspired by its electronic counterpart.

\end{abstract}

\maketitle

Ionic transport is key to numerous processes, from neurotransmission to ultrafiltration~\cite{schoch2008transport,Bocquet:2010kq,Elimelech:2011bsa,Lauger:1985wp}. Over the last decade, it has been extensively investigated in biological systems, evidencing advanced functionalities such as high selectivity, ionic pumping, and electrical or mechanical gating~\cite{Apell:2001eh,kcsa,Dayan2000}. However, it is only recently that experimental progress in nanoscience has allowed fabrication of artificial pores with controlled material properties and (1D or 2D) geometries. These new systems have reached unprecedented nanometre and even angstr\"om scale confinements \cite{Siria:2013dx,Radha:2016en,Feng:2016fh,Tunuguntla:2017js}, yet they are still far from exhibiting the same functions as biological ionic machines. 

At these scales, the ion transport is usually described by the Poisson-Nernst-Planck (PNP) framework~\cite{Bocquet:2010kq}, which couples ion diffusive dynamics to electric interactions. 
Although it may account for non-linear (e.g. diode-type) effects --~\cite{Bocquet:2010kq}, the PNP framework is intrinsically continuum and mean field. 
Building bio-inspired functions, however, may require to control transport at the single ion level, which is out of reach of a mean-field description. Single charge transport actually echoes the canonical Coulomb blockade (CB) phenomenon, which 
 was thoroughly explored in nanoelectronics. 
CB is typically observed in a single electron transistor: under fixed bias, the current between source and drain peaks at quantised values of the gating voltage~\cite{Nazarov}. The origins of this effect stem from the many-body Coulomb interactions between the electrons and from the discreteness of the charge carriers~\cite{Beenakker:1991bb}.
Similar physical ingredients are at play in a nanoscale channel filled with a salt solution (Fig. 1a):  ions also interact via coulombic forces, and a variable surface charge on the channel walls can play the role of the gating voltage. 
One may therefore expect to observe an ``ionic Coulomb blockade'', namely peaks in the ionic conductance of a nanochannel at quantised values of its surface charge. 
It is thus of interest that molecular dynamics simulations~\cite{Stopa:2002gl,Krems:2013jn,Tanaka:2017ik,LiCB} and experiments~\cite{Feng:2016fl,mutagenesis1,mutagenesis2} have shown what might be indirect signatures of ionic CB (though in the absence of a gating voltage), and conductance gating by a surface charge has been demonstrated in simulations of a biological ion channel model~\cite{Kaufman:2012fu,Kaufman:2015fw}.
These observations remain surprising since ionic systems in water at room temperature have specific features contrasting with electronic systems which may preclude the occurrence of ionic CB. Beyond the absence of quantum effects, the fact that ions are of both signs -- while electrons are only negative -- results in Debye screening, which is expected to drastically weaken the many body interaction; it remains unclear under which conditions these aspects may suppress ionic transport quantisation. 

Although pioneering analytical efforts have translated the results established for electrons~\cite{Beenakker:1991bb} to the ionic case~\cite{vonklitzing,Kaufman:2015fw,luchinsky}, a general theory for ionic CB, incorporating the unique features of ionic systems in contrast to their electronic counterparts, is still lacking. 
We develop such a theory in this paper. 

%%%%%%%

\begin{figure*}
\centering
\includegraphics{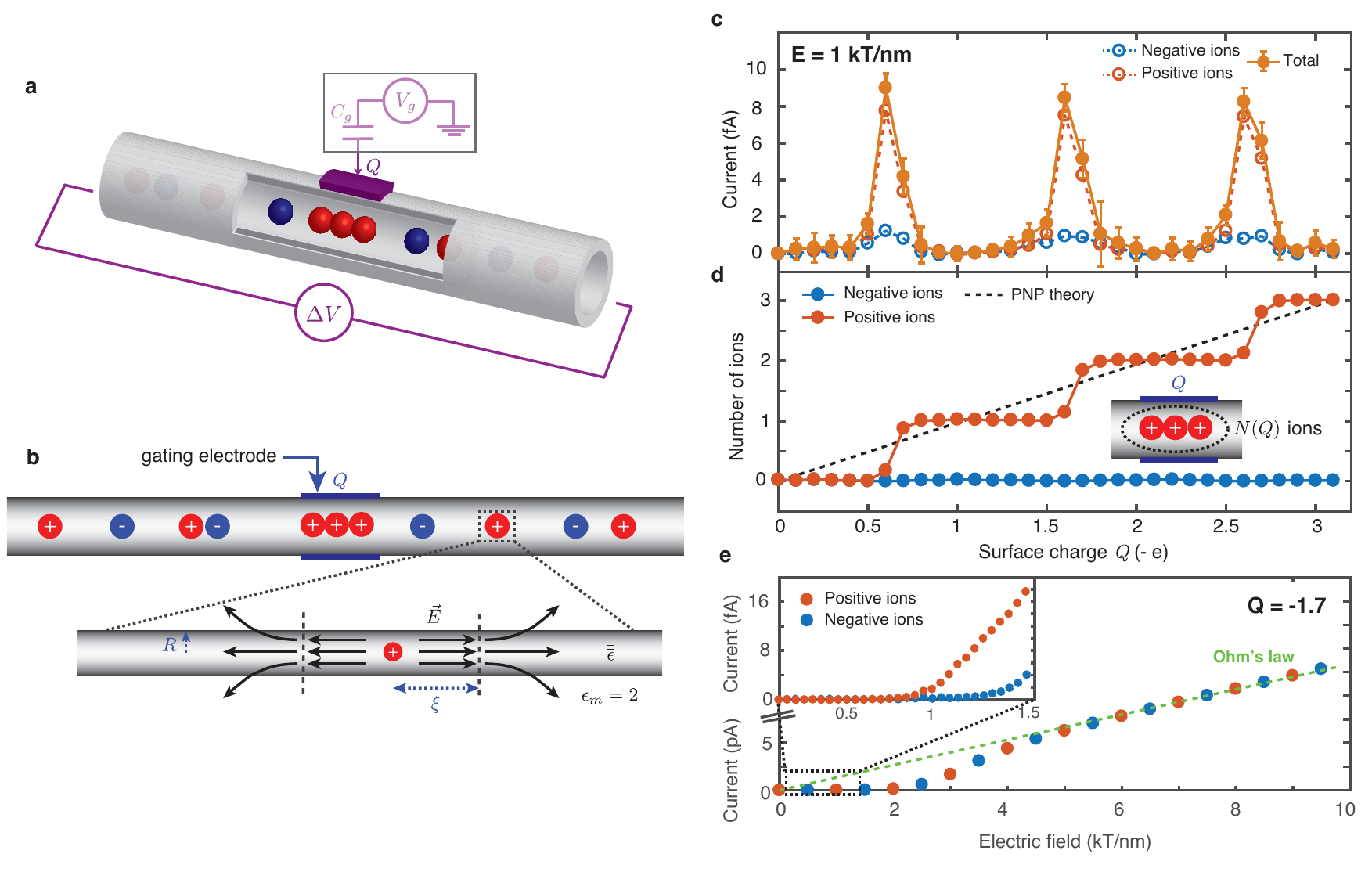}
\caption{\textbf{Brownian dynamics simulations of ionic Coulomb blockade in a nanochannel.} \textbf{a.} Sketch of ions confined in a gated nanochannel. The tuneable surface charge $Q$ acts as a gate; it is equivalent to a voltage applied to an electrode which is capacitively coupled to the channel. \textbf{b.} Setting for the brownian dynamics simulations: the ions have quasi-1D Coulomb interactions, with the electric field confined in the channel over a characteristic length $\xi$.  \textbf{c.} The ionic current through the channel, at fixed applied electric field ($E =1~k_BT/e /\rm nm$), show peaks at discrete values of the surface charge. \textbf{d.} The neutralising charge $N(Q)$ -- the total positive charge that screens the negative surface charge -- increases in steps as a function of $Q$. The dashed line shows the mean-field prediction obtained from the equilibrium solution of the PNP equations (see SI, Sec. 1.5), which is in complete disagreement with the simulations. \textbf{e.} Ionic current as a function of the applied electric field at fixed surface charge ($Q = -1.7$). A strongly non-linear behaviour is observed. The parameters are $x_T = 0.09~\rm nm$, $\xi = 3.5~\rm nm$, and salt concentration $\rho_0 = 0.44~\rm M$. Error bars represent the standard deviation of the mean.}
\end{figure*}

\section*{Model definition and numerical results}

Our theory is based on a simple but general model of a nanochannel which confines ions in one dimension (Fig. 1a). The channel has radius $R$ and length $L \gg R$, as opposed to nanopores of length $L \sim R$. It is filled with water, which under confinement exhibits \emph{a priori} an anisotropic dielectric permittivity $\overline{\overline{\epsilon}}$~\cite{Schlaich:2016dh,fumagalli2018}, and it is embedded in a membrane with low permittivity $\epsilon_m$ (whenever needed, we use $\epsilon_m = 2$). 
Under such conditions (see Fig.~1b), the electric field lines produced by an ion stay confined inside the channel over a characteristic length $\xi$~\cite{Zhang:2005jb}. 
This leads to a stronger Coulomb interaction than in the bulk solution, which is well described by the exponential potential
\begin{equation}
V(x) = k_B T \frac{\xi}{x_T} e^{-|x|/\xi}.
\label{potential}
\end{equation}
This introduces a thermal length $x_T$~\cite{Zhang:2006ee}, which is the typical distance between two opposite charges confined in the channel. We detail in the supplementary information (SI, Sec. 3) how the parameters $\xi$ and $x_T$ are related to the channel geometry and to the various dielectric constants. If the permittivity of confined water is assumed to be the same as in the bulk, one has $\xi \approx 7R$ and $x_T = R^2/2 \ell_B$, where $\ell_B = 0.7~\rm nm$ is the Bjerrum length in bulk water. We shall use these relations in the following, keeping in mind that taking into account the anisotropic permittivity would result in a stronger interaction for a given confinement. 

A charge is imposed on the confining surface and acts as a gate on the system; here
we reduce the surface charge to a point-like charge $Q$. 
The ions interact between themselves and with the surface charge through the potential given in Eq.~\eqref{potential}; depending on conditions, an electric field $E$ may be applied along the channel. 

%%%%%%

Before developing an analytical theory, we confirm using (grand canonical) brownian dynamics simulations~\cite{Cooper:1985cj} that our simplified model displays the ionic CB phenomenology. 
Details of the simulations are given in the SI, Sec. 4: the measured quantities are the ionic current and the neutralising charge $N(Q)$, defined as the total positive charge that screens the negative charge $Q$. Figures 1c and 1d show typical simulation results. Remarkably, we do observe signatures of ionic CB: namely,  the neutralising charge $N(Q)$ is ``quantised'', as it increases in discrete steps as a function of $Q$ -- this can be considered an equilibrium signature of ionic CB --, and under an external electric field, the current peaks at discrete values of $Q$. We thus recover the same phenomenology as in simulations of nanopores~\cite{Kaufman:2012fu,Kaufman:2015fw}, although in our general setting  we do not assume electrostatic coupling between the channel entrances and the surface charge. Furthermore, our simulations reveal a very non-linear current-voltage characteristic (Fig. 1e), with the conductance at low voltages being suppressed with respect to the one expected from Ohm's law. Interestingly, the $I-V$ characteristic for our ionic CB system differs from its electronic counterpart, where several steps in current versus applied voltage are observed before reaching Ohm's law \cite{Nazarov}. 

%%%%%%%%%%%%

\section*{Fractional Wien effect theory}

We now develop an analytical theory in order to understand this counterintuitive behaviour. The vision of Coulomb blockade in terms of energy barriers which has been developed for electrons in quantum dots~\cite{Beenakker:1991bb}, and adapted to ions in nanopores~\cite{Kaufman:2015fw} cannot apply here as we consider a 1D nanochannel of arbitrary length. Moreover, the phenomenon at stake is clearly out of reach for the mean-field PNP equations~\cite{schoch2008transport,Bocquet:2010kq}. 
The PNP result for the neutralising charge is derived in the SI (Sec. 1.5) and is shown in Fig. 1d: a perfectly linear behaviour for $N(Q)$ versus the surface charge $Q$ is obtained, in contrast to the simulation results. Thus, the theoretical description of ionic CB requires to exactly solve the underlying many-body problem. 

To this end, we first compute the grand-canonical partition function of the confined electrolyte in the presence of the gating charge $Q$ and interacting with the pairwise potential of Eq.~\eqref{potential}; the chemical potential is $\mu$ and the temperature $T$ (we set $k_B T = 1$). The system under consideration closely resembles a 1D Coulomb gas model, which can be solved using a functional integral technique as in~\cite{Edwards:1962fj,Demery:2012ey,Kamenev:2006ff}, which we extend to incorporate an arbitrary gating charge density $q(x)$. 
An exhaustive calculation reported in the SI (Sec. 1.2) yields the partition function as

\begin{equation}
\Xi = \int \d\phi_0 \d\phi_L e^{-x_T(\phi_0^2+\phi_L^2)/4\xi} \mathcal{P}(\phi_L,L|\phi_0), 
\label{xi}
\end{equation}
where the propagator $\mathcal{P}(\phi,x|\phi_0)$ solves 
\begin{equation}
\frac{\partial \mathcal{P}}{\partial x} = \frac{1}{x_T} \frac{\partial^2 \mathcal{P}}{\partial \phi^2} + \left(iq\phi - \frac{x_T}{4\xi^2} \phi^2 + \frac{2 z}{L} \cos \phi \right) \mathcal{P}
\label{pde}
\end{equation}
with initial condition $\mathcal{P}(\phi,0|\phi_0) = \delta(\phi-\phi_0)$, and $z = e^{\beta \mu}$ the fugacity. 

This result allows to unveil the unconventional behaviour of the ionic system. 
As a first indication, the equation of state of the confined ionic gas can be exactly derived in the limit $z_T \equiv z\,x_T/L \ll 1$, corresponding to strong electrostatic interactions (with our simulation settings, $z_T = 0.02$), yielding
\begin{equation}
P = \frac{1}{2} \rho k_B T \times \left(1 + O(z_T^2)\right),
\label{state}
\end{equation}
where $P$ is the pressure and $\rho$ is the salt density (see SI Sec.~1.10). Thus, when the interactions are strong enough, the ionic gas of density $\rho$ behaves as an ideal gas of density $\rho/2$. Hence, in the channel, ions of opposite charge are actually bound together in the form of Bjerrum pairs, as confirmed by direct observation in the simulations (supplementary movie S1). The confined salt behaves accordingly as a weak electrolyte. We will now demonstrate that this crucial characteristic, missed by mean-field theory, is the key to explaining both the conductance gating and
the strongly non-linear response under an electric field, as highlighted in Fig. 1.

\begin{figure*}
\centering
\includegraphics{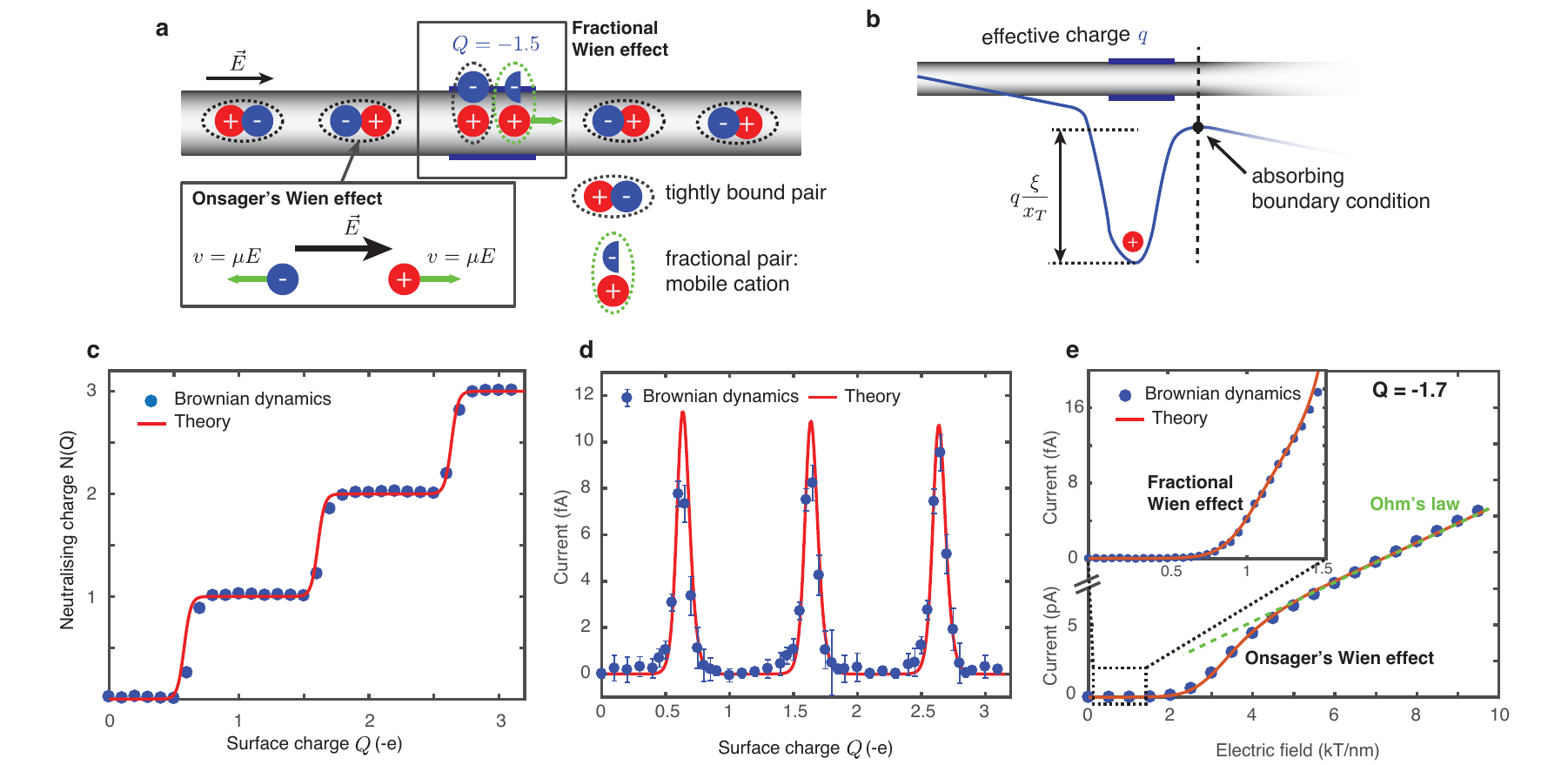}
\caption{\textbf{Analytical theory for the fractional Wien effect mechanism of ionic Coulomb blockade.} \textbf{a.} Schematic representation of Onsager's Wien effect and of the fractional Wien effect. Both correspond to non-linearities in the current-voltage characteristic due to the electric field dependence of the dissociation rate of ion pairs. \textbf{b.} Schematic setting for the study of out-of-equilibrium Bjerrum pair dynamics. The main ingredient is the computation of the mean escape time of an ion from the potential created by an effective charge $q$. \textbf{c.} Neutralising charge $N(Q)$ as a function of surface charge. \textbf{d.} Positive ion current, at fixed applied electric field ($E = 1$ $k_BT/e$/nm), as a function of surface charge. \textbf{e.} Current-voltage characteristic at fixed surface charge, corresponding to a conductance peak ($Q = -1.7$). Inset: zoom on the small applied electric field region. In panels c, d and e, dots are simulations results, while the solid line is the theoretical prediction from Eq.(\ref{Ith}) and Eq.~\eqref{NQ}. 
The Wien effect theory shows quantitative agreement with simulations. Error bars represent the standard deviation of the mean.}
\end{figure*}

Let us first sketch a qualitative picture. In a weak electrolyte, the conduction should proceed through the second Wien effect, which was famously explained by Onsager~\cite{Onsager:2004dv,Kaiser:2013di}. In Onsager's picture, tightly bound ion pairs cannot move under the effect of an electric field, and current can only flow when an ion pair dissociates (Fig.~2a). This picture applies to our system, except for the presence of the gating charge $Q$, which acts as a "defect'' and affects the dissociation process of ion pairs.
If $Q$ is an integer, all the ions are tightly bound at low enough $E$ and the conductance is vanishing. Now if $Q$ has a fractional part, it still binds an integer number of ions, so that one of those ions may be less strongly bound than the others (for example if a charge $Q = -1.5$ binds two positive ions). This weakly bound ion may then dissociate from the surface charge under the effect of the external electric field, resulting in non-zero conductance. This qualitative picture is confirmed by direct observation in the simulations (supplementary movie S1). Interestingly, once a pair dissociates, conduction occurs via a Grotthus-like mechanism, with the free ion exchanging between Bjerrum pairs. 
Altogether, the conduction is due to ions dissociating not from their opposite charge counterparts, but from the fractional surface charge: we thus name this new mechanism  "fractional Wien effect". 

In addition to the surface charge gating, the fractional Wien effect picture allows to understand the non-linear current voltage characteristics. Indeed, at low electric fields, the conduction is due to the dissociation of fractional ion-surface charge pairs, whose dissociation rate depends on the electric field, thus the conductance acquires an electric field dependence. But, at sufficiently high electric fields, the  Bjerrum pairs that are present in the bulk of the channel will also start dissociating, resulting in the standard (``bulk'') Wien effect that was studied by Onsager. 
This results in another non-linearity in the $I-V$ curve, before a collapse onto Ohm's law once all the pairs have been dissociated. 

\begin{figure*}
%\centering
\includegraphics{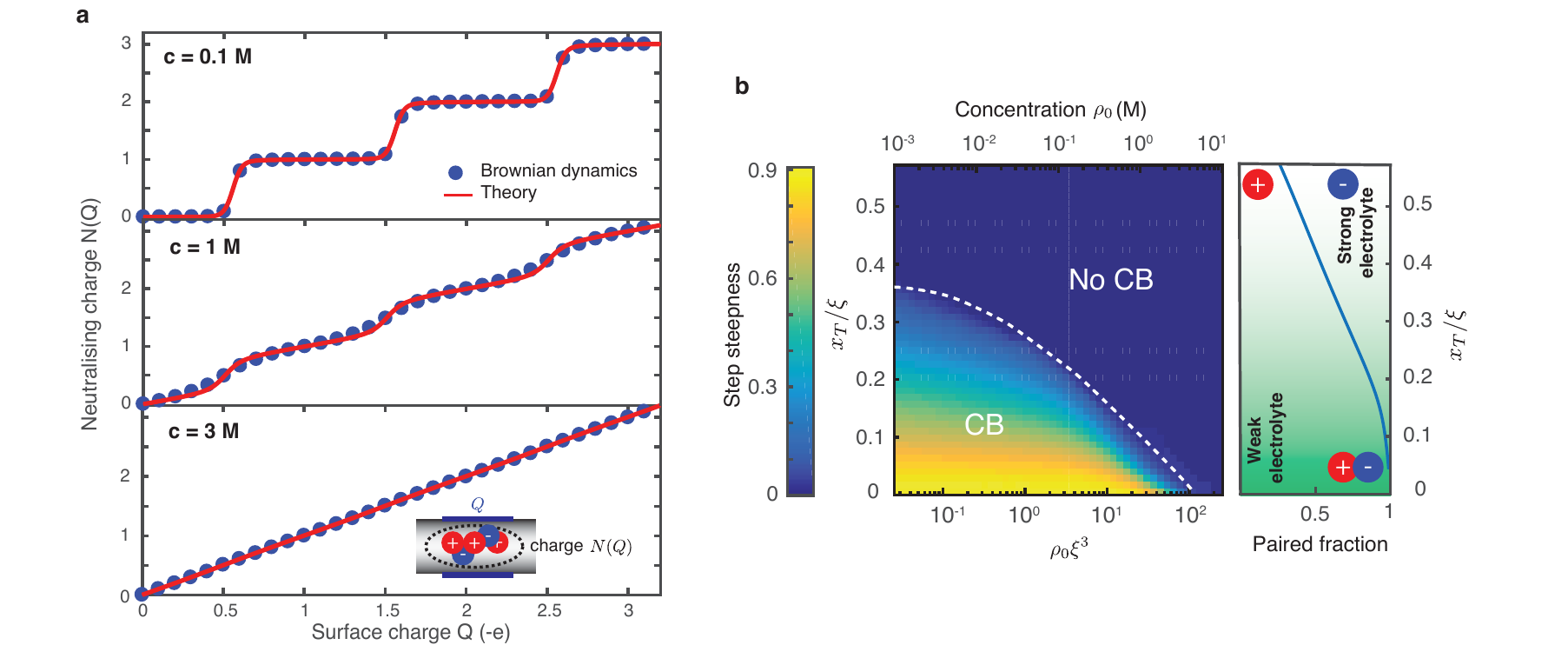}
\caption{\textbf{Conditions for observation of ionic Coulomb blockade.} \textbf{a.} Neutralising charge $N(Q)$ as a function of the surface charge $Q$, as obtained from brownian dynamics simulations (dots), and the full theoretical prediction from Eq.~(S32) (solid line), at different salt concentration values. CB steps disappear as the salt concentration is increased. The parameters are chosen here as $x_T = 0.18~\rm nm$ and $\xi = 7~\rm nm$. \textbf{b.} "Phase diagram" for Coulomb blockade. The colour scale shows, as an assessment of the ``strength'' of ionic CB, the steepness of the steps in the neutralising charge $N(Q)$. It is defined as $1-1/$maximum slope of a step and varies between 0 (no CB) to 1 (full CB). The axes correspond to the channel radius (in terms of the dimensionless quantity $x_T/\xi=R^2/2\ell_B\xi$) and to the dimensionless bulk salt concentration $\rho_0\xi^3$. On the right panel, the fraction of ion pairs in the channel is plotted as a function of $x_T/\xi$, at bulk concentration $0.1~\rm M$. Ion pairing is a prerequisite for ionic Coulomb blockade, {\it i.e.} CB occurs when the salt behaves as a weak electrolyte.}
\end{figure*}

We now develop an out-of-equilibrium framework to quantify the fractional Wien picture that has emerged.
As a first step, the exact solution~\eqref{xi} for the partition function $\Xi$ allows to compute the probability of the system containing a fractional ion-surface charge pair, or equivalently the average neutralising charge $N(Q)$. 
In the limit of an infinite channel and a point-like surface charge, 
we obtain $N(Q) = -(x_T/\xi)( \partial \log \Xi /\partial Q) - Q$ and a lengthy calculation reported in the SI, Sec.~1.9, yields an analytical expression, which can be 
written in the form:
\begin{equation}
N(Q) = \frac{\sum_{ij}a_ia_j (j-i)e^{-\frac{\xi}{2x_T}(Q-(i-j))^2}}{\sum_{ij}a_ia_j e^{-\frac{\xi}{2x_T}(Q-(i-j))^2}}-Q\,;
\label{NQ}
\end{equation}
see SI Eq. (S35) for an exact expression. 
The coefficients $a_i$ are obtained as series expansions in $z_T \equiv z\,x_T/L$ (a few terms are sufficient in the limit of interest $z_T \ll 1$). The prediction in Eq.~\eqref{NQ} is plotted in Fig. 2c: it accounts for a quantised neutralising charge, and the agreement with simulations is excellent.
Assuming strong enough interactions, one may adopt a ``two-state'' perspective: the 
surface charge $Q$ may bind either $\lfloor Q \rfloor$ or $\lfloor Q \rfloor +1$ counterions ($\lfloor . \rfloor$ denoting the floor function); in the latter case, a
fractional ion-surface charge pair is formed. The probability of the system containing this weakly bound pair is accordingly $p(Q) = N(Q) - \lfloor Q \rfloor$. 

In a second step, we study the out-of-equilibrium dynamics of a Bjerrum pair. We consider a single ion bound to an effective charge $q$ (see Fig.~2b). Its probability distribution $P(x,t)$ is governed by the Fokker-Planck equation
\begin{equation} 
\partial_t P = D\partial_x \left( P \partial_x \left[-qV(x)-Ex \right]\right) + D \partial^2_x P,
\label{fp}
\end{equation}
with $D$ the diffusion coefficient, $E$ the applied electric field and $V(x)$ the pairwise interaction potential in Eq.~\eqref{potential}. Solving Eq.~\eqref{fp} with an absorbing boundary condition (see Fig.~2b) yields the mean escape time for the bound ion (see SI Sec. 2.1 and Ref.~\cite{redner}): 
\begin{equation}
 \tau(q,E) = \frac{1}{D} \int_{-\infty}^{+\infty} \int_{\mathrm{max}(0,x)}^{\xi \log \frac{q}{Ex_T}} e^{q(V(x)-V(y))+E(x-y)} \d y \d x .
 \end{equation}
For a bulk ion pair the effective charge $q$ is 1, and the average lifetime of the pair is actually $\tau(1,E)/2$ since both ions are mobile; for a fractional ion-surface charge pair, $q = Q - \lfloor Q \rfloor$. 
In the SI (Sec. 2.2), we derive the relation between the lifetime of the ion pairs and the number of free charge carriers. Combining this result with the probability $p(Q)$ of finding a weakly bound pair in the system yields an expression for the positive ion current $I^+(E)$ accounting for both the fractional and the bulk Wien effect: 
\begin{equation}
I^+(E) = (N(Q)-\lfloor Q \rfloor) I_{Q-\lfloor Q \rfloor}(E) + I^+_{\rm bulk} (E),
\label{Ith}
\end{equation}
with 
$I_q(E) = ({L/(DE)+\tau(q,E)})^{-1}$ and
$I^+_{\rm bulk}(E) = \frac{1}{2\tau(1,E)}\left(\sqrt{1+2\rho D E\tau(1,E)}-1\right)$. 
These analytical predictions for the current (Eq.~\eqref{Ith}), are plotted in Fig. 2d-e: they reproduce 
quantitatively the simulations results. 
Our result accounts both for the CB oscillations -- enhanced conductance at quantised values of surface charge -- 
and for the ``blockade''  of ionic transport in the form of the strongly nonlinear current voltage relation at low applied field. 
The theory fully validates the fractional Wien mechanism, highlighting that this effect originates in an interplay between 
many-body dynamics of ion pairs and Coulomb gas statistics. 

%%%%%%%%

\section*{Phase diagram}

Having now established a theoretical framework for ionic CB that is quantitatively validated against molecular simulations, we may use it to obtain insight into the conditions under which one may expect ionic CB.
Fig. 3a shows the prediction for the neutralising charge $N(Q)$ at increasing salt concentration values: strikingly, the CB steps disappear at high salt concentration as a result of Debye screening, again in full agreement with simulations. 
This is a crucial specificity of our ionic system with respect to its electronic counterpart. 
Going further, we build a phase-diagram which displays the parameter space where ionic CB occurs (that is where $N(Q)$ versus $Q$ displays steps), in terms of dimensionless ion density and channel size, see Fig. 3b. 
Ionic CB indeed disappears above a critical salt concentration for a given channel size (or $x_T$); Debye screening thus does prevent Coulomb blockade, though only at rather high salt concentration values (typically, more than 2 M for a 1 nm channel). Conversely, at a given salt concentration, a small enough $x_T$ (strong enough interactions) is required for ionic CB to occur. 
In the limit $z_T \ll 1$, the slope of a step is given by $(\d N / \d Q)_{\rm max} = \xi/(4x_T) + O(z_T^4)$ (SI Sec. 1.9). Therefore a necessary condition for observing steps is  $\xi/x_T \gtrsim 4$, {\it i.e.} the Coulomb interaction between two ions should be greater than $\sim 4~k_BT$. 
Concretely, this corresponds to nanochannel sizes $R \lesssim 3.5 \ell_B \sim 2.5~\rm nm$ for monovalent ions. 
For multivalent ions with valency $p$, the Bjerrum length increases as $p^2$ and this modifies accordingly the condition on the channel size. Thus, our theory demonstrates that ionic CB can actually be expected in channels that are much larger than previously considered biological nanopores~\cite{Kaufman:2012fu,Kaufman:2015fw} of radius $R \sim 0.3~\rm nm$.  
Finally, the right panel of Fig. 3b shows the fraction of Bjerrum pairs as predicted by our theory (SI Sec. 1.11): it decreases with increasing $x_T$, in line with the disappearance of CB steps, highlighting once more that ionic CB and Bjerrum pairing are intimately related. 

%%%%%%%%%%%

\begin{figure}
\centering
\includegraphics{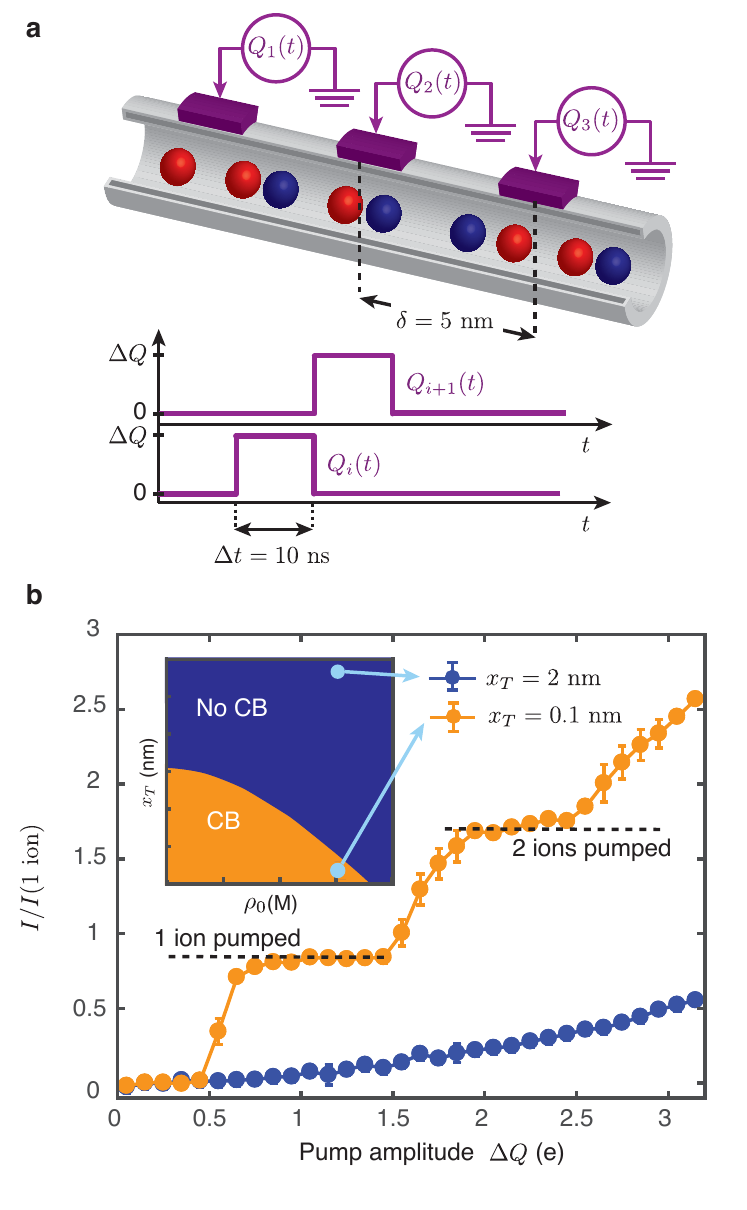}
\caption{\textbf{Coulomb-blockade-based ion pump.} \textbf{a.} Schematic of the ion pump inspired by its electronic counterpart; time dependence of the variable surface charges $Q_i(t)$. The 125 variable charges are placed every 5 nm, and periodic boundary conditions are used. \textbf{b.} Positive ion current resulting from the pump operation (with no applied electric field), at $x_T = 0.1~\rm nm$ (CB regime), and $x_T = 2 ~\rm nm$ (no CB), as obtained from brownian dynamics, as a function of the pump amplitude $\Delta Q$. The current is normalised by $I(1~\rm ion)$, which is the current resulting from one ion moving at a velocity 5 nm/10 ns. Error bars represent the standard deviation of the mean.}
\end{figure}

\section*{Ion pump}

Our modelling of Coulomb blockade opens the way to the design of new functionalities in nanofluidic systems. Beyond gated transport itself, one may also harness the control over single ions allowed by CB to develop ion pumping functionalities. 
Single electron pumps have been obtained by associating two CB devices in series, with their gate voltages oscillating out of phase~\cite{Pothier:1992bx}. Fig. 4a shows an analogous ionic system, a nanochannel with variable surface charges placed along its length. We confirm using brownian dynamics simulations that such a device is capable of pumping activity. Upon appropriate modulation of the surface charges, and in the absence of applied electric field, there is indeed transport of ions along the channel; the modulation amplitude $\Delta Q$ allows to precisely control the discrete number of ions being transported (Fig. 4b). But most importantly, the simulations highlight that the pumping fully relies on the system operating in the CB regime. Indeed, the pumping current is almost 0 at large $x_T$ (weak interactions, no CB), while it is significant only at small $x_T$ (strong interactions, CB regime; see inset of Fig. 4b). 
This proof-of-concept confirms the importance of ionic CB as a building block for artificial devices mimicking biological functions; we leave for future work the thorough investigation of such devices. 

%%%%%%%%%%%%

\section*{Conclusions}

Thanks to a many-body framework for the non-equilibrium ion dynamics -- far beyond the mean field approximation --, we unveil the subtle physical mechanisms underlying the ionic equivalent of Coulomb blockade in a general nanochannel geometry. We highlight that, in stark contrast to its electronic counterpart, ionic CB in a nanochannel relies on the Bjerrum pairing of ions and a fractional Wien-effect mechanism. Building on this fundamental result, we are able to provide a theoretical benchmark for the experimental realisation of ionic CB, establishing prerequisite conditions in terms of confinement and salt concentration, as summarised in the phase diagram in Fig. 3b. 
Ionic CB requires nanometric confinement for monovalent ions but this condition is strongly mitigated for multi-valent species.
These estimates are actually impacted by the  dielectric properties of water in strong confinement: while we propose a way to account for the anisotropic permittivity of confined water~\cite{Schlaich:2016dh,fumagalli2018}, additional insight from simulations or experiments is needed to precisely evaluate the interactions at such scales. Finally, beyond the one-dimensional systems considered here, the recently developed two-dimensional ion channels based on van der Waals heterostructures could also exhibit Coulomb-blockade-like behaviour, with the weaker 2D interactions potentially compensated by the very low permittivity observed in these channels~\cite{fumagalli2018}. 
The exploration of ionic CB in these systems is already within reach, opening exciting perspectives for the development of advanced ionic machinery. 

\section*{Acknowledgements}
The authors thank V. D\'emery, R. Vuilleumier, B. Rotenberg, D. Dean, R. Netz, A. Poggioli, T. Mouterde, L. Jubin and H. Yoshida for useful discussions. SM acknowledges support from ANR Neptune. LB acknowledges support from ERC, project Shadoks and European Union's H2020 Framework Programme/FET {\it NanoPhlow}. This work was granted access to the HPC resources of MesoPSL financed by the Region Ile de France and the project Equip@Meso (reference ANR-10-EQPX-29-01) of the programme Investissements d'Avenir supervised by the Agence Nationale pour la Recherche, as well as  to the HPC resources of CINES under the allocation 2018-A0040710395 made by GENCI. 

\section*{Author contributions}

LB and AS conceived the project. NK carried out the theoretical analysis and brownian dynamics simulations. NK and LB co-wrote the paper. All authors discussed the results and commented on the manuscript. 

\section*{Data availability}
The data that support the plots within this paper and other findings of this study are available from the corresponding authors upon reasonable request.

\section*{Code availability}
The brownian dynamics code used within this study is available from the corresponding authors upon reasonable request.

%\bibliography{cb_ref_complete.bib}

\end{document}

\begin{titlepage}

\vspace*{2cm} \begin{center}
\large Supplementary information for: \par\nobreak
\vspace{20pt}\hrule\vspace{10pt}
\huge\textbf{
Ionic Coulomb blockade \\ as a fractional Wien effect} \par\nobreak
\vspace{10pt}\hrule\vspace{.6cm}
  \vspace{.5cm} 
\large N. Kavokine, S. Marbach, A. Siria and L. Bocquet  \\
\vspace{1cm}
-\\
\vspace{1cm}
\end{center}
\vspace{2cm}
\tableofcontents
\end{titlepage}

\section{Equilibrium properties: Coulomb gas theory}
Our computation is inspired by the original solution of the 1D Coulomb gas model by Lenard and Edwards~\cite{Edwards:1962fj}, and subsequent studies by Demery, Dean and coworkers~\cite{Demery:2016kh,Demery:2012ey,Demery:2012iy,Dean:1998hn}, as well as Shklovskii and coworkers~\cite{Zhang:2006ee,Kamenev:2006ff}.   
\subsection{Model definition}
We consider a one-dimensional lattice with sites $1,\dots,M$ as a model for the nanochannel of radius $R$ and length $L$. Each lattice site $i$ carries a spin $S_i$, which takes the values $\{0,1,-1\}$, corresponding respectively to no ion, a positive ion, or a negative ion occupying the site. We model the surface charge distribution as an extra fixed charge $q_i$ added at each lattice site. The spins interact with the Hamiltonian 
\begin{equation}
\mathcal{H}(\{S_i\}) = \frac{\xi}{2 x_T}\sum_{i,j=1}^M (S_i+q_i) (S_j+q_j) e^{-|i-j|/\xi} \equiv \frac{1}{2x_T} (S+q)^T C (S+q),
\label{hamiltonian}
\end{equation}
where we have taken $k_BT = 1$. The system is in contact with a particle reservoir with bulk density $\rho_0$. Here the parameters $\xi$ and $x_T$ are dimensionless, expressed in number of lattice sites. Their relationship to the geometrical parameters is discussed in section 3 of this Supplementary Information. 

\subsection{General solution}
The grand partition function is given by
\begin{equation}
\Xi = \sum_{S_1,\dots,S_M} z^{\sum_i |S_i|} e^{-\frac{1}{2 x_T} (S+q)^T C (S+q)},
\end{equation}
with $z = \rho_0\pi R^2 L/M$ the fugacity. The matrix $C$ can be analytically inverted: 
\begin{equation}
C^{-1} = \frac{1}{2\xi \sinh (1/\xi)} \cdot 
\left(
\begin{array}{ccccccc}
e^{1/\xi} & -1 & 0 & 0 & \dots & 0 & 0 \\
-1 & 2 \cosh(1/\xi) & -1 & 0 &\dots & 0 & 0 \\
\vdots & \ddots & \ddots & \ddots &  &\vdots &\vdots \\
\vdots &  & \ddots & \ddots & \ddots &\vdots &\vdots \\
\vdots &  &  & \ddots & \ddots &\ddots &\vdots \\
0&0 &\dots &0 & -1& 2\cosh(1/\xi)& -1\\
0& 0&\dots &\dots & 0& -1& e^{1/\xi} \\
\end{array}
\right).
\end{equation}
Hence we can carry out a Hubbard-Stratonovich transformation, that is rewrite the partition function as a gaussian integral, introducing the integration variable $\phi$: 
\begin{equation}
\Xi = \sqrt{\frac{x_T^M}{(2\pi)^M \mathrm{det} (C) }}\cdot \sum_{S_1,\dots,S_M} z^{\sum_i |S_i|} \int \d \phi e^{-\frac{x_T}{2} \phi^T C^{-1} \phi + i(S+q)^T \phi},
\end{equation}
with $\mathrm{det}(C) = \frac{e^{1/\xi}}{2 \sinh (1/\xi)}\cdot \left [ \xi (1-e^{-2/\xi}) \right]^M$. After performing the sum over the spins, which is now decoupled, we obtain
\begin{equation}
\begin{split}
\Xi & =  \sqrt{\frac{x_T^M}{(2\pi)^M \mathrm{det} (C) }}\cdot \int \d \phi_1 \dots \d \phi_M  \prod_{j=1}^M (1 + 2z \cos \phi_j ) \prod_{j=1}^M e^{iq_j \phi_j} \dots \\
& \dots \exp \left(-\frac{x_T}{4 \xi \sinh(1/\xi)}\left[\sum_{j=1}^{M-1} (\phi_{j+1}-\phi_j)^2+ 2(\cosh(1/\xi)-1)\sum_{j=2}^{M-1} \phi_j^2 + (e^{1/\xi} -1)(\phi_1^2 + \phi_M^2)    \right]\right).
\end{split}
\label{HS}
\end{equation}
We now take a continuum limit of the lattice model. We call $a$ the physical lattice spacing and let $\tilde \xi = a \xi$, $\tilde x_T= a x_T$ and $\tilde z = M z$. We then let $a \to 0$ and $M \to \infty$ while keeping the physical length of the system $L = a M$ constant. We then drop the tilde sign to lighten the notation and obtain 
\begin{equation}
\Xi = \int \d \phi(0)  e^{-x_T \phi(0)^2/4\xi} \int [ \d \phi ] e^{-S[\phi]} \int \d \phi(L)e^{-x_T \phi(L)^2/4\xi}
\label{main_pf}
\end{equation}
with 
\begin{equation}
S[\phi] = \int_0^L \d x  \left[\frac{x_T}{4} \left(\frac{\d \phi}{\d x} \right)^2 + \frac{x_T}{4 \xi^2} \phi(x)^2 - i q(x)\phi(x)-\frac{2z}{L} \cos \phi(x) \right] \equiv \int_0^L \mathcal{L}(\phi,\dot \phi).
\label{action}
\end{equation}
$q(x)$ is the one-dimensional density corresponding to the surface charge, and we have changed the notation to $z \equiv \pi R^2 L \rho_0$. At this point $\xi$ and $x_T$ have the dimension of length. The path integral measure is defined as 
\begin{equation}
[\d\phi] = \lim\limits_{\substack{a\to 0 \\ M \to \infty \\ L = aM}} \left[ \prod_{j=1}^M \sqrt{\frac{x_T}{4\pi a}} \d \phi_j \right].
\end{equation}
We now define the propagator $P(\phi,x|\phi_0,0)$, or simply $P(\phi,x)$, as 
\begin{equation}
P(\phi,x) = \int \d \phi(x) \delta(\phi(x)-\phi) \int [\d\phi] e^{-\int_0^x \mathcal{L}(\phi,\dot \phi)} \int \d \phi(0) \delta(\phi(0)-\phi_0).
\end{equation}
Considering an infinitesimal displacement $\Delta x$, 
\begin{equation}
\begin{split}
P(\phi,x) = \sqrt{\frac{x_T}{4\pi \Delta x}} \int \d (\Delta \phi) &P(\phi - \Delta \phi, x- \Delta x) \dots\\
&\dots \exp \left(-\int_{x-\Delta x}^x \d x'  \left[\frac{x_T}{4} \left(\frac{\Delta \phi}{\Delta x} \right)^2 + \frac{x_T}{4 \xi^2} \phi^2 - i q(x)\phi-\frac{2z}{L} \cos \phi \right] \right).
\end{split}
\end{equation}
Expanding the propagator as $P(\phi-\Delta \phi, x - \Delta x) = P(\phi,x) - \Delta x \partial P/\partial x - \Delta \phi \partial P / \partial \phi + (1/2) (\Delta \phi^2) \partial^2P/\partial \phi^2$,   and carrying out the gaussian integrals, we obtain
\begin{equation}
\begin{split}
P(\phi,x) = &\left(P(\phi,x) - \Delta x \frac{\partial P}{\partial x} + O(\Delta x^2) \right) \left( 1 - \Delta x \left[\frac{x_T}{4\xi^2}\phi^2 - i q(x)\phi - \frac{2z}{L}\cos \phi \right] + O(\Delta x^2) \right) \\
& + \frac{\Delta x}{x_T} \frac{\partial^2 P}{\partial x^2}(1+ O(\Delta x)).
\end{split}
\end{equation}
$P(\phi,x)$ thus solves the partial differential equation 
\begin{equation}
\frac{\partial P}{\partial x} = \frac{1}{x_T} \frac{\partial^2 P}{\partial \phi^2} + \left(iq\phi - \frac{x_T}{4\xi^2} \phi^2 + \frac{2 z}{L} \cos \phi \right) P, 
\label{schrod}
\end{equation}
with initial condition $P(\phi,0) = \delta(\phi-\phi_0)$ which is the equivalent of a Schrödinger equation for the path integral representation~\eqref{main_pf}. The partition function can thus be computed as 
\begin{equation}
\Xi = \int \d \phi(L) e^{-x_T \phi^2/4\xi} P(\phi,L|f_0),
\label{pf}
\end{equation}
where $P(\phi,L|f_0)$ is the solution of~\eqref{schrod} with initial condition $P(\phi,0) = f_0 (\phi) \equiv e^{-x_T \phi^2/4\xi}$, which can be easily obtained by finite difference numerical integration.

\subsection{Observables}
In the lattice model, the probability to find, say, a positive ion at position $k$, can be computed by replacing a factor $(1+2z \cos \phi_k) $ by $z e^{i\phi_k}$: 
\begin{equation}
\begin{split}
\langle \rho^+_k \rangle & = \frac{1}{\Xi} \int \d \phi_1 \dots \d \phi_M  \prod_{j\neq k} (1 + 2z \cos \phi_j ) \color{blue}z e^{i\phi_k}\color{black} \prod_{j=1}^M e^{iq \phi_j} \dots \\
& \dots \exp \left(-\frac{x_T}{4 \xi \sinh(1/\xi)}\left[\sum_{j=1}^{M-1} (\phi_{j+1}-\phi_j)^2+ 2(\cosh(1/\xi)-1)\sum_{j=2}^{M-1} \phi_j^2 + (e^{1/\xi} -1)(\phi_1^2 + \phi_M^2)    \right]\right).
\end{split}
\end{equation}
Thus, taking the continuum limit, one can compute the average positive ion density by inserting the operator $z e^{i\phi}$ at position $x$:
\begin{equation}
\langle \rho^+(x) \rangle = \frac{1}{\Xi}\int \d \phi(0)\d\phi(x) \d \phi(L)  e^{-x_T \phi(0)^2/4\xi} P(\phi(x),x|\phi(0),0) \color{blue}z e^{i \phi(x)}\color{black} P(\phi(L),L|\phi(x),x) e^{-x_T \phi(L)^2/4\xi},
\end{equation}
which can again be obtained by finite differences.
Let us now compute the electrostatic potential $\Phi$ at a lattice point $j_0$. One has 
\begin{equation}
\begin{split}
\frac{\partial{\Xi}}{\partial q_{j_0}} &= \sum_{S_1, \dots, S_M}\left[-\frac{1}{x_T}\sum_i  (S_i + q_i) C_{ij_0} \right] z^{\sum_i |S_i | } e^{- \mathcal{H}(\{S_i \})} \\
&= \sum_{S_1, \dots ,S_M} \left[ - \frac{\xi}{x_T} \sum_i (S_i + q_i) e^{-|i-j_0|}\right] z^{\sum_i |S_i | } e^{- \mathcal{H}(\{S_i \})} \equiv - \langle \Phi_{j_0} \rangle \cdot \Xi
\end{split}
\end{equation}
Now looking at eq.~\eqref{HS}, differentiating $\Xi$ with respect to $q_{j_0}$ corresponds to inserting a factor $i\phi_{j_0}$, thus in the continuum limit the electrostatic potential is computed by inserting an operator $-i \phi$: 
\begin{equation}
\langle \Phi(x) \rangle = \frac{1}{\Xi}\int \d \phi(0)\d\phi(x) \d \phi(L)  e^{-x_T \phi(0)^2/4\xi} P(\phi(x),x|\phi(0),0)\color{blue}(-i \phi(x))\color{black}P(\phi(L),L|\phi(x),x) e^{-x_T \phi(L)^2/4\xi}.
\end{equation}

\subsection{Effect of ion valence}
So far we have only considered monovalent ions. Ions of valence $p$ could be taken into account by having the spins $S_i$ take the values $\{ p,0,-p\}$ instead of $\{1,0,-1\}$. The action~\eqref{action} then becomes 
\begin{equation}
S[\phi] = \int_0^L \d x  \left[\frac{x_T}{4} \left(\frac{\d \phi}{\d x} \right)^2 + \frac{x_T}{4 \xi^2} \phi(x)^2 - i q(x)\phi(x)-\frac{2z}{L} \cos p \phi(x) \right],
\end{equation}
or, after a change of variable,
\begin{equation}
S[\phi] = \int_0^L \d x  \left[\frac{(x_T/p^2)}{4} \left(\frac{\d \phi}{\d x} \right)^2 + \frac{(x_T/q^2)}{4 \xi^2} \phi(x)^2 - i (q(x)/p)\phi(x)-\frac{2z}{L} \cos  \phi(x) \right].
\end{equation}
Thus any property for ions of valence $p$ will be the same as for ions with valence 1, with $p^2$ times stronger interactions and $p$ times smaller surface charge. 

\subsection{Mean-field approximation}
The mean-field version of the Coulomb gas model is obtained by taking a saddle point approximation in path integral~\eqref{main_pf}. Minimising the action~\eqref{action} with respect to the function $\phi(x)$, we obtain a differential equation for $\phi(x)$: 
\begin{equation}
\frac{x_T}{2} \frac{\d^2 \phi}{\d x^2} = \frac{x_T}{2\xi^2} \phi - i q + \frac{2 z}{L} \sin \phi .
\end{equation}
Following the argument in section 1.3, we identify the dimensionless electrical potential as $\Phi = - i \phi$. Thus $\Phi$ satisfies 
\begin{equation}
\left(  \frac{\d^2}{\d x^2} - \frac{1}{\xi^2} \right)\Phi = - \frac{2q}{x_T} + \frac{4 z}{L x_T} \sinh \Phi.
\end{equation}
Now using $x_T = R^2/(2\ell_B) = 2\pi \epsilon_0\epsilon_w R^2/e^2$ (we still take $k_BT = 1$) and $z = \rho_0 \pi R^2 L$, we obtain 
\begin{equation}
\left(\frac{\d^2}{\d x^2}-\frac{1}{\xi^2}\right)\Phi(x) = -\frac{q(x)}{\epsilon_w \epsilon_0} + \frac{2\rho_0}{\epsilon_w \epsilon_0} \sinh \Phi(x).
\label{PB}
\end{equation}
This the equivalent of the Poisson-Boltzmann equation (that is, the Poisson-Nernst-Planck equations at equilibrium) for our system. The dashed black line in Fig. 1d of the main text corresponds to the numerical solution of eq.~\eqref{PB}. It could have also been derived in the standard way from the Boltzmann distribution of ions in the electrostatic potential $\Phi$, which solves the modified Poisson equation $(\d^2/\d x^2 - 1/\xi^2)\Phi = -\rho_c/(\epsilon_w\epsilon_0)$, with $\rho_c$ the charge density. The Green's function of this modified Poisson's equation is precisely the pairwise potential in eq. (1) of the main text. 

\subsection{Self-energy}
An ion confined in the channel creates an electric field $\mathbf{E}(x)$ such that $|E(x)| = e^{-|x|/\xi}/x_T$. Thus, it has an electrostatic self-energy
\begin{equation}
E_s = \pi R^2 \int \d x \frac{\epsilon}{2} E(x)^2 = \pi R^2 \frac{\epsilon}{2 x_T^2} \xi = \frac{\xi}{2 x_T}. 
\end{equation}
One can check that this self-energy is taken into account in the hamiltonian~\eqref{hamiltonian}: each particle contributes $\frac{\xi}{2 x_T}$ to the system's energy. 

\subsection{Fourier space and thermodynamic limit}
In practice, equation~\eqref{schrod}, is most easily solved in Fourier space. We define
\begin{equation}
\tilde P (k,x) = \frac{1}{\sqrt{2\pi}} \int \d \phi e^{-ik\phi} P(\phi,x).
\end{equation}
Then $\tilde P(k,x)$ satisfies
\begin{equation}
\frac{\partial \tilde P}{\partial x} = - \frac{k^2}{x_T} \tilde P - q \frac{\partial \tilde P}{\partial k} + \frac{x_T}{4 \xi^2}\frac{\partial^2\tilde P}{\partial k^2} + \frac{z}{L} \left[\tilde P(k+1,x)+\tilde P(k-1,x) \right],
\end{equation}
or, defining $\tilde x \equiv x/x_T$ (and $\tilde L \equiv L/x_T$),
\begin{equation}
\frac{\partial \tilde P}{\partial \tilde x} = - k^2 \tilde P - qx_T \frac{\partial \tilde P}{\partial k} + \frac{x_T^2}{4 \xi^2}\frac{\partial^2\tilde P}{\partial k^2} + z_T \left[\tilde P(k+1,\tilde x)+\tilde P(k-1,\tilde x) \right].
\label{kspace}
\end{equation}
$qx_T$ is the number of surface charges per length $x_T$, and $z_T = \rho_0 \pi R^2 x_T$ is the number of ions per length $x_T$ if the density in the channel was the same as in the reservoir. 

The term proportional to $\partial \tilde P / \partial k$ in eq.~\eqref{kspace} is an advection term: it induces a drift in $k$ space, by an amount which equals the total surface charge $Q$. We assume from now on that the distribution $q(x)$ is reduced to a point: $q(x) = Q \delta(x-L/2)$, or, after the change of variable, $q(\tilde x) = (Q/x_T)\delta(\tilde x - \tilde L/2)$. Then, the effect of the advection term is an infinitely fast shift by an amount $Q$, that is the action of an operator
\begin{equation}
S_Q : f \mapsto (g:k\mapsto f(k-Q)).
\end{equation}
Let us also define the operator $\mathcal{T}$ such that 
\begin{equation}
[\mathcal{T}(\tilde P)] (k) = - k^2 \tilde P + \frac{x_T^2}{4 \xi^2}\frac{\partial^2\tilde P}{\partial k^2} + z_T \left[\tilde P(k+1,\tilde x)+\tilde P(k-1,\tilde x) \right],
\end{equation}
which plays the role of a functional transfer matrix. Recalling eq.~\eqref{pf}, the partition function then reads
\begin{equation}
\Xi = \langle f_0 | e^{\frac{\tilde L}{2}\mathcal{ T}}S_Q e^{\frac{\tilde L}{2}\mathcal{T}} | f_0 \rangle 
\end{equation}
with $f_0(k) = e^{-\xi k^2/x_T}$ and $\langle f(k) | g(k) \rangle \equiv \int \d k f^*(k) g(k)$. 

Now in the limit $L \to \infty$, we may consider the largest eigenvalue $\lambda$ of the operator $\mathcal{T}$, and the associated eigenfunction $\chi$: 
\begin{equation}
[\mathcal{T} ( \chi)] (k) = \lambda \chi (k).
\end{equation}
Then, up to an exponentially small correction,
\begin{equation}
\Xi = |\langle f_0 | \chi \rangle |^2 \langle \chi(k) | \chi(k-Q) \rangle e^{\lambda \tilde L}.
\label{pcharge}
\end{equation}

\subsection{Observables in the thermodynamic limit}
The insertion of  $e^{i\phi}$ in direct space is the equivalent of a unit shift in Fourier space. Hence the ion density at a point $\tilde x$ (assuming for simplicity $\tilde x < \tilde L/2$) is given by 
\begin{equation}
\langle \rho_{\pm} (\tilde x) \rangle = z_T\frac{ \langle \chi(k-Q) | [e^{- (\tilde L/2-\tilde x) \mathcal{T}} \cdot \chi(k\mp1)] \rangle}{ \langle \chi(k-Q) | [e^{- (\tilde L/2-\tilde x) \mathcal{T}} \cdot \chi(k)] \rangle},
\label{rhopm}
\end{equation}
in ions per length $x_T$. The function $\chi$ can be computed by finite difference integration of eq.~\eqref{kspace}: in practice, we start from the initial condition $\tilde P(k,0) = f_0(k)$ and carry out the integration until convergence to the eigenfunction.  Further numerical integration allows to compute $[e^{- (\tilde L/2-\tilde x) \mathcal{T}} \cdot \chi(k)] $. We define the neutralising charge $N(Q)$ as the average charge in the interval $[L/2-1.25~\mathrm{nm};L/2+1.25~\mathrm{nm}]$, which can thus be computed by integration of the density obtained from eq.~\eqref{rhopm}. The plots of $N(Q)$ versus $Q$ obtained in such a way were used to obtain the data in figure 3. 
In an uncharged channel (or far away from the surface charge), eq.~\eqref{rhopm} is reduced to 
\begin{equation}
\langle \rho_{\pm}^{\mathrm{bulk}} \rangle = z_T \frac{\langle \chi(k) | \chi(k\mp1)\rangle}{\langle \chi(k) | \chi(k)\rangle},
\label{rhobulk}
\end{equation}

Insertion of $-i\phi(x)$ in direct space corresponds to differentiating with respect to $q(x)$ in Fourier space. Hence, in particular, the potential at $x = L/2$ can be computed as
\begin{equation}
\langle\Phi (L/2) \rangle = -\frac{\langle \chi(k) | \chi'(k-Q)\rangle}{\langle \chi(k) | \chi(k-Q)\rangle}.
\label{potential}
\end{equation}
In the Coulomb blockade regime, the $N(Q)$ ions are closely bound to the surface charge. Since an ion creates in its immediate vicinity a potential $\xi/x_T$, 
$[N(Q)+Q](\xi/x_T) = \langle\Phi(L/2)\rangle,$
and therefore the neutralising charge is
\begin{equation}
N(Q) =  -Q-  \frac{x_T}{\xi}\frac{\langle\chi(k)|\chi'(k-Q)\rangle}{\langle\chi(k)|\chi(k-Q)\rangle}.
\label{NQ}
\end{equation}

\subsection{Strong coupling approximation}
\begin{figure}
\centering
\includegraphics{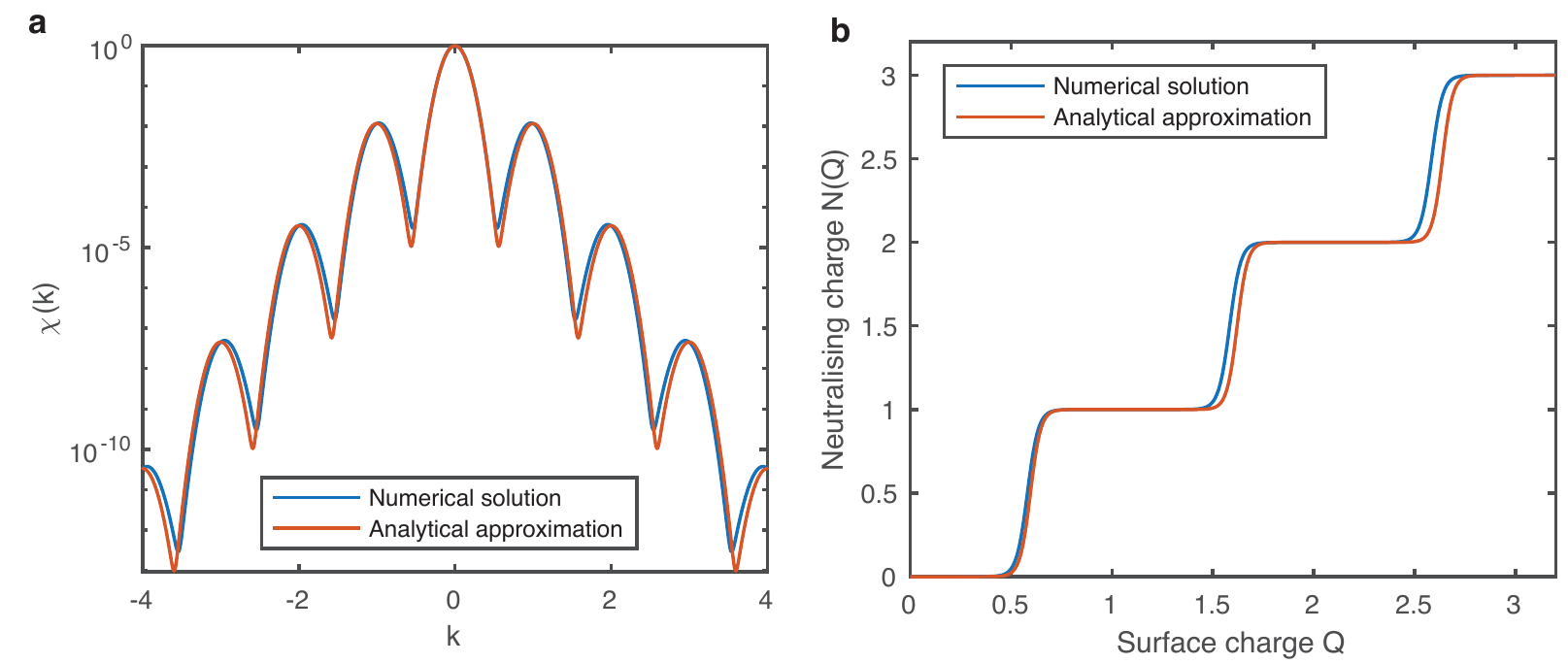}
\caption{\textbf{a.} Function $\chi(k)$, as computed by finite differences from equation \eqref{kspace}, and analytically from equation \eqref{chik}, with $z_T = 0.012$ ($x_T = 0.9 \AA$ and $\rho_0 = 0.28~\rm M$). \textbf{b.} Neutralising charge $N(Q)$, as computed by finite differences from equation \eqref{rhopm}, and analytically from equation \eqref{NQan}, with $z_T = 0.012$ ($x_T = 0.9 \AA$ and $\rho_0 = 0.28~\rm M$).}
\label{an_num}
\end{figure}

Strong interactions correspond to low values of $x_T$ and, if $\rho_0$ is not too large, then the condition $z_T \ll 1$ holds. For instance, in our simulations, $x_T = 0.09~\rm nm$, and for a channel of radius $R = 0.5~\rm nm$, $z_T = 0.042 \times \rho_0$ (in M). Thus, for describing our system in the Coulomb blockade regime, it is reasonable to consider only the first terms in the expansion of observables in powers of $z_T$. To obtain such an expansion, we first need to compute the function $\chi$. Equation~\eqref{kspace} without the diffusion term, \begin{equation}
\frac{\partial \tilde P}{\partial \tilde x} = - k^2 \tilde P + z_T \left[\tilde P(k+1,\tilde x)+\tilde P(k-1,\tilde x) \right],
\label{mathieu}
\end{equation}
is actually the Fourier-transformed Mathieu equation, and methods for computing the Fourier coefficients of its highest periodic eigenfunction $\chi_0$ are known from the mathematical literature~\cite{AS} (see Appendix): 
\begin{equation}
\chi_0(k) = \frac{1}{2} \left[ a_0 \delta(k) +\sum_{i > 0} a_i (\delta (k-i)+\delta(k+i)) \right]
\end{equation}
with 
\begin{align}
&a_0 = 2  \\
\label{a0}
&a_{1} = 2z_T - \frac{7}{2} z_T^3 + \frac{116}{9}z_T^5+ O(z_T^7)\\
&a_{2} = \frac{1}{2} z_T^2 - \frac{10}{9}z_T^4 +O(z_T^6)\\
&a_{3} = \frac{1}{18} z_T^3 - \frac{13}{96}z_T^5+ O(z_T^7)\\
&a_{4} = \frac{1}{288} z_T^4 + O(z_T^6)\\
\label{a3}
&a_{ p > 4}  = O(z_T^p) 
\end{align}
The function $\chi_0(k)$ is represented by an ensemble of discrete peaks at integer values of $k$. The effect of the diffusion term in eq.~\eqref{kspace} will be to spread these discrete peaks over a non-zero width. In practice, keeping the diffusion term and taking $z_T = 0$, equation~\eqref{kspace} becomes 
\begin{equation}
\frac{\partial \tilde P}{\partial \tilde x} = - k^2 \tilde P + \frac{x_T^2}{4 \xi^2}\frac{\partial^2\tilde P}{\partial k^2},
\end{equation}
which is solved by 
\begin{equation}
\tilde P(k,\tilde x) = e^{-\frac{x_T^2}{4 \xi^2}\tilde x}e^{-\xi k^2/x_T}.
\label{solz0}
\end{equation}
Thus one may heuristically assume that if the width $\sqrt{x_T/\xi}$ of solution~\eqref{solz0} is much smaller than 1, an eigenfunction of the operator $\mathcal{T}$ will be well approximated by a set gaussian peaks centred on integer values of $k$. The highest eigenfunction $\chi$ will then be given by
\begin{equation}
\chi(k) = e^{-\xi k^2/x_T}+ \frac{1}{2}\sum_{i\in \mathbb{Z^*}} a_{|i|} e^{ -\xi(k-i)^2/x_T}.
\label{chik}
\end{equation}
This analytical approximation is actually in good agreement with the numerical result obtain from finite differences, as shown in figure S\ref{an_num}. 

We can now use equation \eqref{NQ} to obtain an analytical expression for the neutralising charge. After computing the gaussian integrals, 
\begin{equation}
\boxed{N(Q) = \frac{\sum_{ij}a_ia_j (j-i)e^{-\frac{\xi}{2x_T}(Q-(i-j))^2}}{\sum_{ij}a_ia_j e^{-\frac{\xi}{2x_T}(Q-(i-j))^2}}-Q}~.
\label{NQan}
\end{equation}

Figure S\ref{an_num} shows the neutralising charge obtained by numerical integration from equation \eqref{rhopm} and the analytical result from equation \eqref{NQan}. They are in very good agreement, especially for $Q \lesssim 1$, since for larger values of $Q$ higher orders in $z_T$ are involved. The first step ($N(Q), Q \leq 1$) has the simple form
\begin{equation}
N_1(Q) = \frac{1}{1+\frac{1}{2z_T}e^{-\frac{\xi}{x_T}\left( Q - \frac{1}{2}\right)}}.
\end{equation}
The maximum slope is attained at 
\begin{equation}
Q_{\rm max} = \frac{1}{2} - \frac{x_T}{\xi} \log(2z_T)
\label{qmax}
\end{equation}
and the corresponding slope is 
\begin{equation}
\left.\frac{\d N}{\d Q} \right|_{Q_{\rm max}} = \frac{\xi}{4 x_T} + O(z_T^4),
\end{equation}
so that steps are apparent when $x_T \geq \xi/4$ as stated in the main text. Equation \eqref{qmax} shows that the steps are shifted to the right with respect to the half-integer values of $Q$ by an amount that varies as the logarithm of the salt concentration. We thus recover on rigorous grounds the logarithmic dependence that was suggested in ref.~\cite{Kaufman:2017ci}. 

\subsection{Equation of state}
The pressure of the Coulomb gas is defined as 
\begin{equation}
P = - \frac{\partial F}{\partial L},
\end{equation}
with $F$ the free energy. The total number $N$ of ions can be obtained by deriving the free energy with respect to the chemical potential $\mu$ (which is the same for positive and negative ions). Given that $z = e^{\mu}$, 
\begin{equation}
N = - z \frac{\partial F}{\partial z}.
\end{equation}
Thus, the density is related to the pressure by
\begin{equation}
\rho = \frac{\partial N}{\partial L} = z \frac{\partial P}{\partial z}.
\end{equation}
Now the total ion density can be computed from eq.~\eqref{rhobulk}, and using eq.~\eqref{chik} for the function $\chi(k)$ in the strong coupling approximation, we obtain
\begin{equation}
\rho = 2 z_T  \frac{\langle \chi(k) | \chi(k-1)\rangle}{\langle \chi(k) | \chi(k)\rangle} =  2z_T\frac{\sum_{ij}a_{|i-1|}a_{|j|} e^{-\frac{\xi}{2x_T}(i-j)^2}}{\sum_{ij}a_{|i|}a_{|j|} e^{-\frac{\xi}{2x_T}(i-j)^2}},
\end{equation}
the lengths being now expressed in units of $x_T$. The above sums can be considerably simplified if we may neglect the terms with $i\neq j$. The order $p$ in $z_T$ to which such an approximation is valid is given by 
\begin{equation}
z_T^p \sim e^{-\frac{\xi}{2x_T}} , ~\text{i.e.} ~ p \sim \frac{\xi}{2x_T\log(1/z_T)}.
\end{equation}
With the parameter values corresponding to our simulations, we obtain $p \sim 4.4$. We therefore neglect the terms with $i\neq j$ and expand the coefficients $a_i$ up to order $z_T^4$. We obtain: 
\begin{equation}
\rho = 4 z_T^2 - 14 z_T^4 + O(z_T^6)
\end{equation}
and 
\begin{equation}
\boxed{P = \frac{\rho}{2} \left(1+\frac{7}{4}z_T^4 + O(z_T^6)\right)}~.
\end{equation}
Thus up to order $z_T^4$ we obtain the same equation of state as was derived for the ideal Coulomb gas ($\xi \to \infty$) in ref.~\cite{Dean:1998hn}. It demonstrates that, in the strong coupling limit, the Coulomb gas of density $\rho$ behaves as an ideal gas of density $\rho/2$: the positive and negative ions are associated in tightly bound pairs, that is the salt behaves as a weak electrolyte. 

\subsection{Paired fraction}
Since the ions in the Coulomb gas associate in tightly bound pairs, we may be interested in estimating the fraction of paired ions for a given set of parameters. We may only give a qualitative estimate, since it depends on the way we define an ion pair. Consider a negative test charge placed at $x = x_0$ in a homogeneous Coulomb gas ($Q=0$). If it is in a paired state, then there is a positive charge that we assume to be uniformly distributed in an interval $[x_0-x_T;x_0+x_T]$. (At this point, we may have chosen any interval $[x_0-\lambda x_T,x_0+\lambda x_T]$, with $\lambda$ of order 1, which would give a quantitatively different result). The test charge therefore feels a potential
\begin{equation}
\Phi_{\rm paired} = \frac{1}{x_T} \int_0^{x_T} \frac{\xi}{x_T}e^{-x/\xi} = \left( \frac{\xi}{x_T} \right)^2 (1- e^{-x_T/\xi}) \equiv \frac{\xi}{x_T} f(x_T).
\end{equation}
If the test charge is unpaired, then it feels a potential $\Phi_{\rm unpaired}$ created by a negative charge uniformly distributed in an interval $[x_0-\delta,x_0+\delta]$, with $\delta = 1/(2\rho^-)$, $\rho^-$ being the negative ion density: 
\begin{equation}
\Phi_{\rm unpaired} = \frac{\xi}{x_T} f(\delta)
\end{equation}
Now the average potential at the location of the test charge can be computed from eq.~\eqref{potential}: 
\begin{equation}
\Phi(x_0) = - \frac{\langle \chi(k) | \chi'(k+1)\rangle}{\langle\chi(k)|\chi(k+1)\rangle}.
\end{equation}
From the above considerations, if we denote $n_p$ the fraction of paired ions, this potential may also be expressed as
\begin{equation}
\Phi(x_0) = \frac{\xi}{x_T}(n_p f(x_T) + (1-n_p)f(\delta)-1).
\end{equation}
We thus obtain the paired fraction as 
\begin{equation}
n_p = \frac{((x_T/\xi)\Phi(0)-1)-f(\delta)}{f(x_T)-f(\delta)}.
\label{np}
\end{equation}
Eq.~\eqref{np} was used to compute the data in figure 3b of the main text. 

\subsection*{Appendix}
In this appendix, we provide a method for deriving the series expansions of the coefficients $a_i$ (eqs.~\eqref{a0}--\eqref{a3}), following ref.~\cite{AS}. We are looking for a solution of eq.~\eqref{mathieu}, 
\begin{equation}
\frac{\partial \tilde P}{\partial \tilde x} = - k^2 \tilde P + z_T \left[\tilde P(k+1,\tilde x)+\tilde P(k-1,\tilde x) \right],
\end{equation}
in the form 
\begin{equation}
\tilde P(k,\tilde x) = e^{\lambda\tilde  x} \sum_{i \geq 0} \frac{1}{2}a_{i} (\delta (k-i)+\delta(k+i)).
\end{equation}
Substituting this ansatz into eq.~\eqref{mathieu} we find the set of equations 
\begin{equation}
\left\{
\begin{array}{l}
\lambda a_0/2 - z_Ta_1 = 0\\
-(\lambda+1)a_1+z_Ta_0+z_Ta_2 = 0\\
-(\lambda+k^2)a_k + z_T(a_{k+1}+a_{k-1}) = 0, \forall k >1
\end{array}
\right.
\label{system}
\end{equation}
We now define  $\Lambda_k \equiv (\lambda+k^2)/z_T$. Successively rearranging the equations of system~\eqref{system}, we obtain
\begin{equation}
\Lambda_0 = \frac{2a_1}{a_0} = \frac{2}{\Lambda_1-\dfrac{a_2}{a_1}} = \frac{2}{\Lambda_1-\dfrac{1}{\Lambda_2-\dfrac{a_3}{a_2} }} = ...
\end{equation}
This yields a closed equation for the $\Lambda_k$'s in the form of a continued fraction: 
\begin{equation}
0 = \Lambda_0- \frac{2}{\Lambda_1-\dfrac{1}{\Lambda_2- \dfrac{1}{\Lambda_3 - \dfrac{1}{\Lambda_4 -...}} }}.
\label{contfrac}
\end{equation}
We can now perturbatively solve eq.~\eqref{contfrac} in the limit of small $z_T$. Starting at order 2, we have to solve $\Lambda_0-2/\Lambda_1 = 0$, which yields two solutions for $\lambda$: $\lambda_1 = 2 z_T^2 + O(z_T^4)$ and $\lambda_2 = -1 - 2z_T^2 + O(z_T^4)$. Since we are looking for the largest eigenvalue $\lambda$ we pick only the first solution. Having fixed the coefficient at order 2, we obtain only one solution for $\lambda$ at the next orders. Iterating the procedure, we obtain the expansion
\begin{equation}
\lambda = 2z_T^2 - \frac{7}{2}z_T^4+\frac{116}{9}z_T^6- \frac{68687}{1152}z_T^8 + \frac{123707}{400} z_T^{10} + O(z_T^{12}),
\end{equation}
in agreement with ref.~\cite{Dean:1998hn}. Knowing the expansion of $\lambda$ to arbitrary order, we can solve the system~\eqref{system} truncated at the $\rm n^{th}$ equation to obtain the coefficients $a_i$ to the $\rm n^{th}$ order in $z_T$, yielding eqs.~\eqref{a0}--\eqref{a3}.

\section{Bjerrum pair dynamics and fractional Wien effect}
The aim of this section is to provide a theory of non-equilibrium ion transport that accounts for the ionic Coulomb blockade phenomenology. From the equilibrium Coulomb gas theory and from the simulations we know that in the Coulomb blockade regime the salt behaves as a weak electrolyte, where the positive and negative ions are strongly correlated as they from tightly bound pairs. However, it is reasonable to assume that the globally neutral ion pairs are largely uncorrelated. Hence we will study the dynamics of a single ion pair, and then incorporate the statistics of the Coulomb gas in order to describe the fractional Wien effect. \emph{We recall that we use units such that $k_B T=1$ and $e=1$.}

\subsection{Lifetime of an ion pair}
We consider a mobile ion of charge $+1$, with diffusion coefficient $D$ in one dimension, interacting with a fixed charge $-q$ placed at $x = 0$, under the effect of an electric field $E$. The two point charges interact with the pairwise potential given by eq. (1) of the main text: $V(x) = \frac{\xi}{x_T} e^{-|x|/\xi}$. We wish to compute the average time $\tau (q,E)$ the mobile ion takes to escape from the confining potential $-qV(x)-Ex$, if placed initially at $x = 0$. We follow the well-known method described for example in~\cite{redner}. The probability distribution $P(x,t)$ of the mobile ion is governed by the Fokker-Planck equation
\begin{equation}
\partial_t P = D\partial_x \left( P \partial_x \left[-qV(x)-Ex \right]\right) + D \partial^2_x P,
\label{fp}
\end{equation}
with initial condition $P(x,0) = \delta(x)$. We consider the ion has escaped once it has crossed the potential barrier on the right, the crossing point $\delta$ being defined by 
$\frac{\d}{\d x}\left. (qV(x) -Ex) \right|_{x=\delta} = 0$, that is 
\begin{equation}
\delta(q,E) = \xi \log \left( \frac{q}{E x_T} \right).
\end{equation}
Therefore we place an absorbing boundary condition at $x = \delta$ as schematically depicted in figure 2b of the main text. 

The probability that the ion has already escaped at time $t$ is $1-\int  P(x,t)\d x$. Thus, the probability density for escaping at time $t$ is $-\int  \partial_t P(x,t) \d x$, so the average time it takes to escape is 
\begin{equation}
\tau(q,E) = \int_0^{\infty} t \left( -\int_{-\infty}^{+\infty} \partial_t P(x,t) \d x\right) \d t,
\end{equation}
or, after integration by parts, 
\begin{equation}
\tau(q,E) =  \int_{-\infty}^{+\infty} \d x \int_0^{\infty} \d t   P(x,t) .
\end{equation}
We now define the time-integrated probability density $G(x) = \int_0^{\infty} P(x,t) \d t$, which solves the time-integrated version of eq.~\eqref{fp}: 
\begin{equation}
-\delta(x) = D\partial_x \left( G \partial_x \left[-qV(x)-Ex \right]\right) + D \partial^2_x G.
\label{intfp}
\end{equation}
We solve eq.~\eqref{intfp} enforcing: 
\begin{itemize}
\item{The absorbing boundary condition $G(\delta) = 0$.}
\item $G(x \to -\infty) = 0$. 
\item The continuity of $G$ at $x = 0$, $G(0+) = G(0-)$. 
\item The discontinuity of $\partial_x G$ as imposed by the $\delta$ function: $\partial_x G|_{0^+}-\partial_x G|_{0^-} -qG(0)(\partial_x V|_{0^+}-\partial_x V|_{0^-}) = -1/D$. 
\end{itemize}
We find a solution 
\begin{equation}
G(x) = \frac{1}{D} \int_{\mathrm{max}(0,x)}^{\delta} e^{q(V(x)-V(y))+E(x-y)} \d y, 
\end{equation}
hence the mean escape time 
\begin{equation}
\tau(q,E) = \int_{-\infty}^{+\infty} G(x) \d x = \frac{1}{D} \int_{-\infty}^{+\infty} \int_{\mathrm{max}(0,x)}^{\xi \log \frac{q}{Ex_T}} e^{q(V(x)-V(y))+E(x-y)} \d y \d x. 
\end{equation}
If we are considering a mobile ion coupled to an effective surface charge $q$, then the average lifetime of such a pair is indeed $\tau(q,E)$. But our computation also applies to a bulk ion pair: one should take $q = 1$ and replace the diffusion coefficient $D$ by $2D$ since both ions are mobile. Therefore the average lifetime of a bulk ion pair is $\tau(1,E)/2$. 

\subsection{From pair lifetime to ionic current}
We shall separately consider the current resulting from ions dissociating from the fractional surface charge and from bulk ion pairs breaking apart. 

The contribution from the surface charge dominates the ionic current at low electric fields, when bulk ion pairs do not dissociate. When all the bulk ions are paired, an ion that dissociates from the surface charge moves unhindered at a velocity $DE$, dragged by the electric field, through the periodic boundary condition, until it comes back to bind to the surface charge, after a time $t_f = L/(DE)$. We assume that the Grotthus-like exchange does not significantly affect $t_f$. Thus, an ion bound to an effective surface charge $q$ is actually free for a fraction $t_f/(t_f +\tau(q,E))$ of the time, during which it contributes $I_0 = 1/t_f$ to the ionic current, so the resulting average current is
\begin{equation}
I_q(E) = \frac{t_f}{t_f + \tau(q,E)} I_0. 
\end{equation}

At larger electric fields, when the bulk ion pairs start to dissociate, the contribution of the single ion that may be released by the surface charge becomes negligible. Let $N$ be the total number of positive ions and $N_f$ the average number of free positive ions. When a bulk pair dissociates, the two free ions which are released travel in opposite directions, each producing a current $I_0$, until they encounter a free ion of the opposite sign, which occurs on average after a time $t_f/(2N_f)$. Thus $N_f$ solves the self-consistent equation
\begin{equation}
\frac{N_f}{N} = \frac{t_f/N_f}{t_f/N_f + \tau(1,E)},
\label{selfc}
\end{equation}
from which we find $N_f$ and the resulting positive ion current, 
\begin{equation}
I^+_{\rm bulk}(E) = N_f I_0= \frac{t_f}{2\tau(1,E)}\left(\sqrt{1+\frac{4N\tau(1,E)}{t_f}}-1\right)I_0. 
\end{equation}
\subsection{Coulomb gas statistics and effective charge}
We now discuss the value of the effective charge $q$. In the Coulomb blockade regime a surface charge $Q$ can bind either $\lfloor Q \rfloor$ counterions with probability $1-p(Q)$ or $\lfloor Q \rfloor +1$ counterions with probability $p(Q)$. In the first case, all the counterions feel an effective charge which is larger than 1, so that the contribution to the current from their dissociation is negligible with respect to the contribution of bulk ion pairs. In the second case, however, there is a counterion which feels a charge $q = Q-\lfloor Q \rfloor < 1$ and may therefore have a non-negligible contribution to the current. The average number of counterions is $N(Q) = (1-p(Q))\lfloor Q \rfloor + p(Q) (\lfloor Q \rfloor +1)$, hence $p(Q) = N(Q)-\lfloor Q \rfloor$, with $N(Q)$ known from the equilibrium Coulomb gas theory (eq.~\eqref{NQan}). If we admit that the current values can be averaged with the grand canonical probabilities of the corresponding numbers of counterions, we obtain an expression for the total positive ion current in the whole electric field range: 
\begin{equation}
I^+(E) = (N(Q)-\lfloor Q \rfloor) I_{Q-\lfloor Q \rfloor}(E) + I^+_{\rm bulk} (E).
\end{equation}
The result given by this equation is plotted in figure 2 of the main text. 

\subsection{Discussion}
The first theory of the Wien effect was famously established by Onsager in 1934~\cite{Onsager:2004dv}. However, it applies to the case of a three-dimensional electrolyte, interacting with a $1/r$ Coulomb potential. With our one-dimensional electrolyte and effective potential $V(x)$, we could not exactly follow Onsager's approach. Onsager computes the normalised two-point function $g(r) = \langle \rho_+(r) \rho_-(0) \rangle /( \langle \rho_+ \rangle \langle \rho_- \rangle )$, which solves the Fokker-Planck equation 
\begin{equation}
\mathbf{\nabla} \cdot (g\mathbb{\nabla} [-V(\mathbf{r}) - Ex]) + \mathbf{\nabla}^2 g = 0. 
\label{fponsager}
\end{equation}
If $V(\mathbf{r})$ is a real Coulomb potential, it solves $\nabla^2 V = 0$, and thus a constant function is a solution of~\eqref{fponsager}. Solutions of~\eqref{fponsager} can then be decomposed as a sum of a constant function, which corresponds to dissociated ion pairs, and a function which goes to 0 at infinity, which corresponds to tightly bound pairs. Onsager shows that the probability currents corresponding to these two solutions give access respectively to the association and dissociation rates of ion pairs. However, our one-dimensional effective potential $V(x)$ does not solve $\partial^2_x V = 0$, hence no such decomposition is possible. The 1D solution of~\eqref{fponsager} satisfying $g(\pm \infty) = 1$ is 
\begin{equation}
g^E(x) = E e^{V(x)+Ex}\int_x^{+\infty} e^{-V(y)-Ey}\d y,
\end{equation}
but it does not give a straightforward access to the average number of ion pairs. One may resort to the drastic approximation used by Liu~\cite{liu,kaiser:thesis}, which amounts to assimilating a ratio of integrals of $g$ to a ratio of values at $x = 0$ at small electric fields. In this approximation the relative increase in the number $N_f$ of free ions when applying an electric field $E$ is given by
\begin{equation}
\frac{N_f(E)}{N_f(0)} = \sqrt{\frac{g^0(0)}{g^E(0)}} = \left(E \int_0^{+\infty} e^{-V(y)-Ey}\d y\right)^{-1/2}.
\end{equation}
Unfortunately this result does not agree well with simulations, nor with our escape rate theory. 

Even if the technical difficulties related to an Onsager-type approach could be overcome, it would not take into account correlations between ion pairs. While these correlations are negligible in three dimensions, they are important in one dimension: two unpaired ions of opposite sign will necessarily encounter each other. The escape rate approach, on the contrary, does allow to take such correlations into account: for instance, they are introduced by eq.~\eqref{selfc}. In three dimensions, escape rate approaches have been attempted~\cite{mcilroy}, but they could only proceed through strong approximations, and the end results thus poorly compared with Onsager's theory and with simulations~\cite{kaiser:thesis}. 

\section{Coulomb interaction in confinement}

In this section we derive the expression of the pairwise interaction potential in eq. (1) of the main text, and relate its parameters $\xi$ and $x_T$ to the radius $R$ of the channel. The exponential form of eq. (1), $V(x) = \frac{\xi}{x_T} e^{-\xi/x_T}$, accounts for a quasi-1D Coulomb interaction. It has been proposed in 2005 by Teber~\cite{teber}, for a model system where a nanochannel of radius $R$ (and length $L \gg R$) is filled with water of high dielectric permittivity, and is embedded in a membrane of much lower permittivity. Teber assumes for the confined water the same dielectric permittivity as for the bulk water, $\epsilon_w = 80$. We know, however, that this assumption may not hold: in strong confinement, the dielectric permittivity of water becomes anisotropic and its value $\epr$ in the confined direction may be very different from its value $\epp$ along the direction that is not confined~\cite{Schlaich:2016dh,fumagalli2018}. Here we compute exactly the electrostatic potential in the anisotropic permittivity case and discuss the effect of the anisotropy on the parameters $\xi$ and $x_T$~\footnote{The computation is inspired by Cihan Ayaz, ``Tensorial Electrostatics in Planar and Cylindrical Geometry'', Bachelorarbeit with Roland Netz,  Freie Universit\"at Berlin, Fachbereich Physik (2016)}. 

We consider the nanochannel of radius $R$ in cylindrical coordinates $(r,z)$. We denote  $\overline{\overline \epsilon}$ the dielectric permittivity tensor. We place a point charge $q$ at $(r,z)=(0,0)$ and solve the Poisson equation for the electrostatic potential $\Phi$, which reads
\begin{equation}
\nabla \left( \overline{\overline \epsilon} \cdot \nabla \Phi \right) = -\frac{q}{\epsilon_0} \frac{\delta(r) \delta(z)}{2 \pi r}.
\label{poisson}
\end{equation}
The tensor $\overline{\overline \epsilon} $ is diagonal in cylindrical coordiantes,
\begin{equation}
\overline{\overline \epsilon} = 
\left(
\begin{array}{ccc}
\epsilon_r & 0 & 0 \\
0 & \epsilon_{\theta} & 0 \\
0 & 0 & \epsilon_z \\
\end{array}
\right).
\end{equation}
Inside the channel ($r < R$), the permittivity is that of water and therefore $\epsilon_r = \epsilon_{\theta} = \epr$ and $\epsilon_z = \epp$. Outside the channel, we assume an isotropic permittivity $\epsilon_m$, so that $\epsilon_r = \epsilon_{\theta} = \epsilon_z = \epsilon_m$. 

We may expand eq.~\eqref{poisson} as
\begin{equation}
\epsilon_r \partial^2_r \Phi + \frac{\epsilon_r}{r} \partial_r \Phi + \epsilon_z \partial^2_z \Phi = -\frac{q}{\epsilon_0} \frac{\delta(r) \delta(z)}{2 \pi r}.
\label{poisson_exp}
\end{equation}
We now define the Fourier-transformed potential $\tilde \Phi (r,k)$ by 
\begin{equation}
\Phi(r,z) = \frac{1}{2\pi^2} \int_0^{\infty} \d k \, \tilde \Phi(k,r) \cos(kz), 
\label{fourier_cos}
\end{equation}
and eq.~\eqref{poisson_exp} in Fourier space reads 
\begin{equation}
 \partial^2_r\tilde \Phi + \frac{1}{r} \partial_r \tilde \Phi -  \left(k\sqrt{\frac{\epsilon_z}{\epsilon_r}}\right)^2 \tilde \Phi = -\frac{q}{\epsilon_0\epsilon_r} \frac{\delta(r) }{r}.
 \label{poisson_fourier}
\end{equation}
We look for solutions of~\eqref{poisson_fourier}
separately in the regions $r < R$ and $r >R$. The solutions are given by: 
\begin{equation}
\begin{array}{rl}
r < R: & \tilde \Phi_< (k,r) = A I_0 \left(kr\sqrt{\epp/\epr}\right) + B K_0 \left(kr\sqrt{\epp/\epr}\right) \\
r > R: & \tilde \Phi_> (k,r) = C I_0 (k r) + DK_0 (k r) \\
\end{array}
\label{solbessel}
\end{equation}
where $A,B,C,D$ are four constants and $I_0$ and $K_0$ are the modified Bessel functions. Since $I_0$ diverges at infinity, we immediately set $C = 0$. We now determine the boundary conditions at $r = 0$ and $r = R$. 
The continuity of the potential at $r = R$ implies 
\begin{equation}
A I_0 \left(kR\sqrt{\epp/\epr}\right) + B K_0 \left(kR\sqrt{\epp/\epr}\right) = DK_0(kR).
\label{cont}
\end{equation}
To determine the condition on the electric field at $r = 0$, we integrate eq.~\eqref{poisson_fourier} over an interval $[r =-\lambda; r=\lambda]$: 
\begin{equation}
\int_{-\lambda}^{\lambda} \d r \, (r \partial^2_r + \partial_r)\tilde \Phi_<(k,r) - \left(k \sqrt{\epp/\epr}\right)^2 \int_{-\lambda}^{\lambda} \d r \, r \tilde \Phi_<(k,r) = - \frac{q}{\epsilon_0 \epr}\int_{-\lambda}^{\lambda} \delta(r) \d r.
\end{equation}
Recalling that $(r \partial^2_r + \partial_r) = \partial_r(r \partial_r)$ and letting $\lambda$ go to 0, we obtain the condition 
\begin{equation}
\lim_{r \to 0} \left[ r \partial_r \tilde \Phi_< (k,r)\right] = - \frac{q}{\epsilon_0\epr}.
\label{r0}
\end{equation}
Applying the same procedure at $r = R$, we obtain 
\begin{equation}
\epr\left. \partial_r\tilde \Phi_< \right|_{r = R} = \epsilon_m \left. \partial_r\tilde \Phi_> \right|_{r = R}.
\label{rR}
\end{equation}
In terms of the solution~\eqref{solbessel}, the condition~\eqref{r0} reads
\begin{equation}
k \sqrt{\frac{\epp}{\epr}} \cdot B \cdot \lim_{r \to 0} \left[r K_1 \left(kr\sqrt{\frac{\epp}{\epr}}\right)\right] = \frac{q}{\epsilon_0\epr}, 
\label{cond0}
\end{equation}
and since $K_1 (x) \underset{x\to 0}{\sim} 1/x$, this yields 
\begin{equation}
B = \frac{q}{\epsilon_0 \epr}.
\label{Bsol}
\end{equation}
Similarly, the condition~\eqref{rR} implies
\begin{equation}
\sqrt{\epr \epp}\cdot \left[A I_1 \left(kR\sqrt{\epp/\epr}\right) - B K_1\left(kR\sqrt{\epp/\epr}\right) \right] = -\epsilon_m D K_1(kR).
\label{condR}
\end{equation}
Solving together eqs.~\eqref{cont}, \eqref{Bsol} and \eqref{condR} yields the constants $A,B,C$ and the expression for the Fourier-transformed potential inside the channel: 
\begin{equation}
\begin{split}
\tilde \Phi_<(k,r) & = \frac{\sqrt{\epr \epp} K_1 \left(kR\sqrt{\epp/\epr}\right) K_0(kR) - \epsilon_m K_1(kR) K_0 \left(kR\sqrt{\epp/\epr}\right)}{\sqrt{\epr \epp} K_0 \left(kR\sqrt{\epp/\epr}\right)I_1(kR) + \epsilon_m K_1 \left(kR\sqrt{\epp/\epr}\right)I_0(kR)} \cdot \frac{q I_0 \left(kr\sqrt{\epp/\epr}\right) }{\epsilon_0 \epr}\dots \\
&  + \frac{q}{\epsilon_0 \epr} \cdot K_0  \left(kr\sqrt{\epp/\epr}\right) .
\end{split}
\label{result_potential}
\end{equation}
Setting $\epp = \epr$ one recovers eq. (2) of ref.~\cite{teber}. We may now follow ref.~\cite{teber} in identifying the first term in eq.~\eqref{result_potential} as the contribution of the images charges in the confining medium and the second term as the usual $1/r$ Coulomb potential. Indeed, upon reverse Fourier transformation, it yields a contribution 
\begin{equation}
\Phi_0 (r,z) = \frac{q}{4 \pi \epsilon_0 \sqrt{(z\epr)^2+(r \sqrt{\epp \epr})^2}},
\end{equation}
which dominates the potential $\Phi(r,z)$ at short distances $r \ll R$. Thus, interestingly, it is the radial permittivity that matters for the short-range interaction along the longitudinal direction.

Now at distances $r \gg R$, one needs only to consider the contribution of small wavelengths $k$ (such that $kR \ll 1$) in eq.~\eqref{fourier_cos}. We set $\kappa \equiv kR$ and expand the denominator of eq.~\eqref{result_potential} at small $\kappa$. We obtain: 
\begin{equation}
\begin{split}
&\sqrt{\epr \epp} K_0 \left(\kappa\sqrt{\epp/\epr}\right)I_1(\kappa) + \epsilon_m K_1 \left(\kappa\sqrt{\epp/\epr}\right)I_0(\kappa) =\\
&\frac{\epsilon_m}{\kappa} \sqrt{\frac{\epr}{\epp}} \left[1+ \frac{\kappa^2}{4 \epsilon_m \epr}\left( \epsilon_m(\epr-\epp) + 2 \gamma \epp (\epsilon_m - \epr) + 2 \epp (\epsilon_m - \epr)\log \left( \frac{\kappa}{2} \sqrt{\frac{\epp}{\epr}} \right) \right)+o(\kappa^2) \right],
\end{split} 
\label{expansion}
\end{equation}
where $\gamma$ is Euler's gamma constant. In order to be able to compute the reverse Fourier transform, Teber proposes to introduce a characteristic length $\xi$ such that the term in brackets in eq.~\eqref{expansion} may be simplified as $1 + (\kappa\xi/R)^2$. Enforcing that the bracket should equal 2 when $\kappa = R/\xi$, one obtains an implicit equation for $\xi$, which reads in our anisotropic permittivity case 
\begin{equation}
\frac{\xi^2}{R^2} = \frac{1}{4 \epsilon_m\epr} \left[\epsilon_m(\epr-\epp)+2\epp(\epsilon_m-\epr)\left(\gamma - \log\left(\frac{2\xi}{R}\sqrt{\frac{\epr}{\epp}}\right) \right) \right].
\label{xinew}
\end{equation}
Setting $\epr = \epp = \epsilon_w$ and assuming $\epsilon _m \ll \epsilon_w$ , one recovers 
\begin{equation}
\frac{\xi^2}{R^2} = \frac{\epsilon_w}{2\epsilon_m} \left[ \log \left( \frac{2\xi}{R} - \gamma \right) \right],
\end{equation}
which is eq. (7) of ref.~\cite{teber}. Provided that the above approximation scheme holds in the anisotropic permittivity case, one obtains the potential along the $z$ direction as 
\begin{equation}
\Phi(0,z) = \Phi_0(0,z) + \frac{q}{2\pi^2 \epsilon_0 R} \frac{\sqrt{\epp}}{\epsilon_m \epr \sqrt{\epr}} \int_0^{+\infty} \d \kappa \, \kappa \cdot \frac{\sqrt{\epr \epp} K_1 \left(\kappa\sqrt{\epp/\epr}\right)K_0(\kappa) - \epsilon_m K_0 \left(\kappa\sqrt{\epp/\epr}\right)K_1(\kappa) }{1+(\kappa\xi/R)^2}.
\end{equation}
Teber identifies the length $\xi$ as the point of transition between a 1D-like linear Coulomb potential and a $1/r$ tail. He thereby approximates the potential as 
\begin{equation}
\Phi(0,z) = \frac{\xi}{x_T} e^{-|z|/\xi},
\end{equation}
where $x_T \approx R^2/(2 \ell_B)$, with $\ell_B$ is the Bjerrum length for an isotropic water permittivity $\epsilon_w$. In the case of a not too strongly anisotropic permittivity, one may use our eq.~\eqref{xinew} to compute a corrected value of $\xi$. However, if the permittivity is strongly anisotropic there is no solution to eq.~\eqref{xinew}. Hence one may not use the exponential approximation directly: qualitatively, it is the short-range $1/r$ term in the potential that dominates over the 1D term that comes from the image charges. 
\begin{figure}
\centering
\includegraphics[scale=0.9]{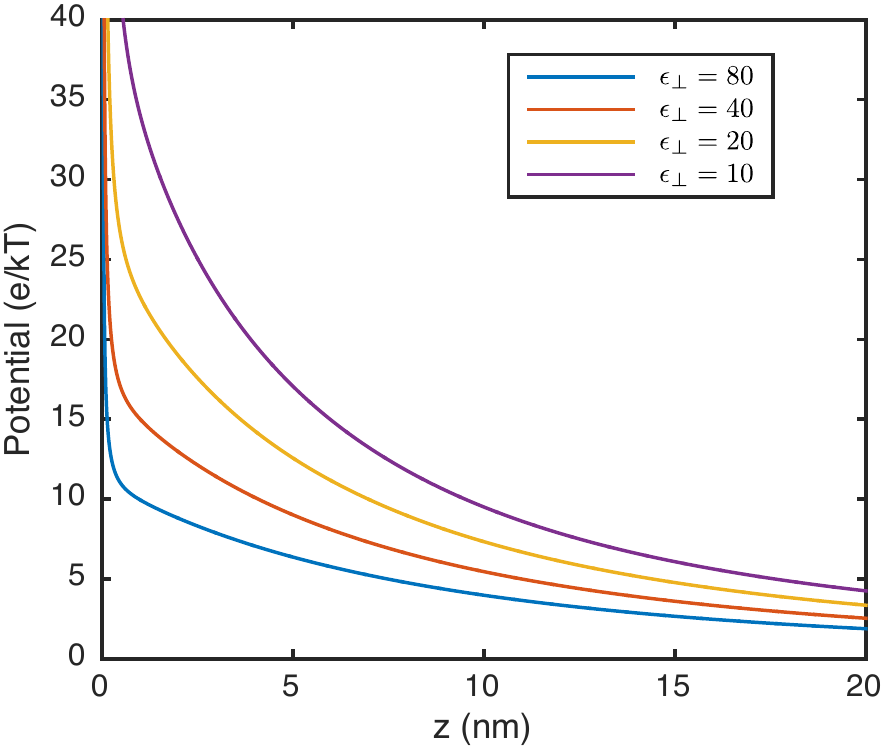}
\caption{Electrostatic potential along the channel axis $z$ as computed from eq~\eqref{result_potential}, for $R = 1~\rm nm$, $\epsilon_m = 2$, $\epp = 80$ and a range of values of $\epr$.}
\label{fpotential}
\end{figure}
One may then use the exact expression~\eqref{result_potential} to estimate the parameters $\xi$ and $x_T$ that would suitably describe the shape of the potential. Figure S\ref{fpotential} shows the exact result for the potential $\Phi(z,0)$, for $R = 1~\rm nm$, $\epsilon_m = 2$, $\epp = 80$ and a range of values of $\epr$. Qualitatively, the potential along the channel axis is stronger for lower values of $\epr$, thus we always underestimate the Coulomb interaction strength by setting the dielectric permittivity of water to its bulk value. 

\section{Simulation methods}
\subsection{Brownian dynamics}
Our 1D brownian dynamics simulations are carried out using the LAMMPS software~\cite{lammps}. The simulation system consists of $N_+$ positive ions and $N_-$ negative ions, in a one-dimensional box of length $L$ with periodic boundary conditions. An immobile point charge $Q$ is placed at $x = L/2$ to model the surface charge. We typically use $N_- = 100$, $N_+ = 100 + n$ with $0 \leq n \leq 3$, and $L = 12.5~\mu \rm m$, unless stated otherwise. We only simulate the motion of ions (the solvent is implicit), and the ion positions at timestep $i+1$ are determined from the positions at timestep $i$ by solving a Euler-discretised overdamped Langevin equation: 
\begin{equation}
x_{i+1} = x_{i} - \Delta t \frac{e D}{k_B T} \partial_x \Phi |_{x=x_i}+ \eta_i\sqrt{2D \Delta t},                                                                                                    
\end{equation}
where $\Phi$ is the electrostatic potential and $\eta_i$ a gaussian random variable of 0 mean and unit variance. We use a timestep $\Delta t = 5~\rm ps$, diffusion coefficient $D = 10^{-9}~\rm m^2\cdot s^{-1}$ and temperature $T = 298~\rm K$. The potential $\Phi$ acting on the ion $i$ takes into account the interaction with the ions $j \neq i$ with the pairwise potential $V(x) = \frac{\xi}{x_T} e^{-|x|/\xi}$, the interaction with the surface charge $Q$ with the same pairwise potential, and the contribution $-Ex$ from the applied electric field $E$. Unless stated otherwise, we used the values $x_T = 0.9~ \AA$ and $\xi = 3.5~\rm nm$, which physically correspond to divalent ions confined in a channel of diameter $1~\rm nm$. With these values of the parameters, the ion density we imposed in the channel (200 ions per 12.5 $\mu$m) corresponds to a salt concentration $c = 0.44~\rm M$ in the reservoirs, as determined from eq.~\eqref{rhobulk}. In each simulation run, we measured the neutralising charge $N(Q)$ as the average number of positive ions in the interval $[L/2 - 1.25~\mathrm{nm}, L/2 + 1.25~\mathrm{nm}]$, and the positive and negative ion currents: 
\begin{equation}
I_{\pm} = \left\langle\frac{\pm e}{L}\sum_{\pm ~ \rm ions} \frac{x_{i+1}-x_i}{\Delta t} \right \rangle_{\rm dynamics}
\end{equation} 
The ions were initially randomly distributed in the simulation box. The simulations lasted for $5\times10^8$ timesteps, with the first $10^7$ timesteps left for equilibration. Error bars represent the standard deviation of the sampled observable, corrected by its correlation time. 

\begin{figure}
\centering
\includegraphics{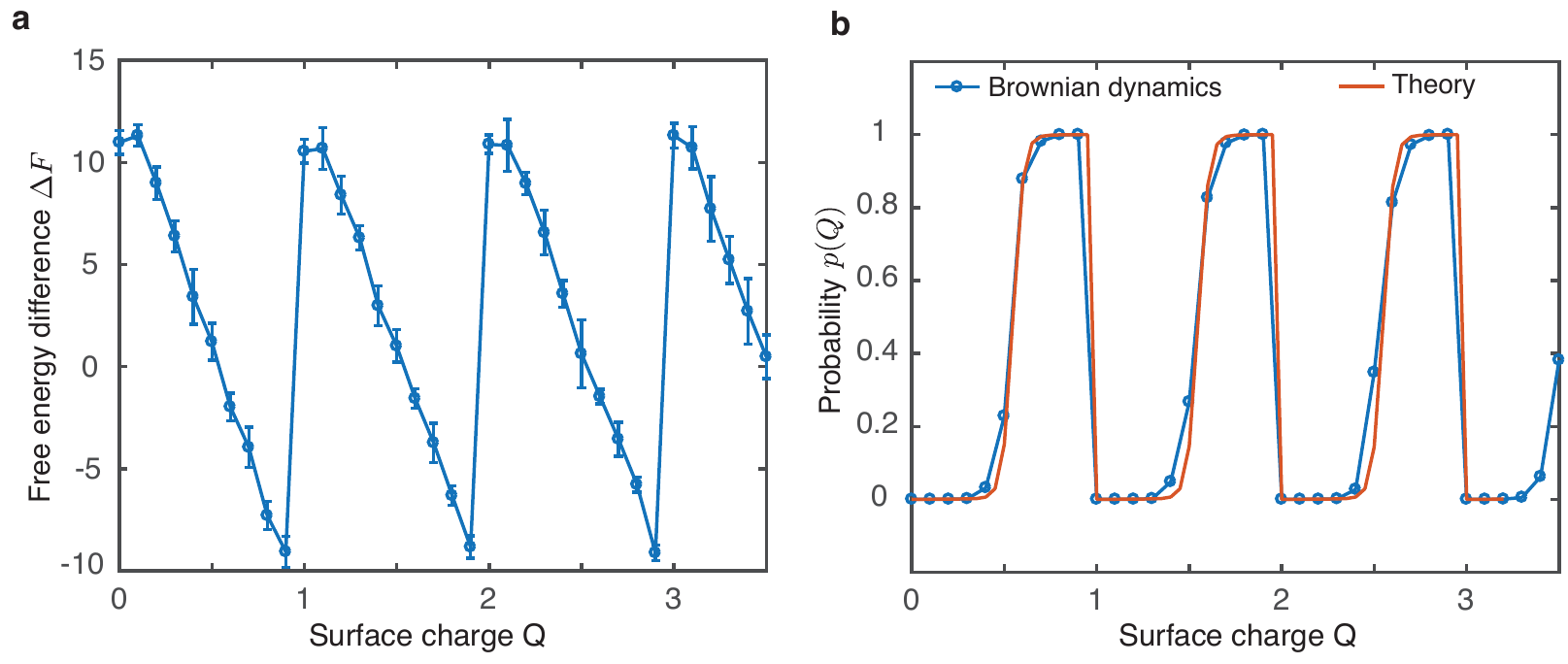}
\caption{\textbf{a.} Free energy difference $\Delta F = F(\lfloor Q \rfloor +1)-F(\lfloor Q \rfloor)$, as a function of surface charge $Q$, as obtained from the brownian dynamics simulations by the thermodynamic integration method. \textbf{b.} Probability $p(Q)$ of the channel containing $N_+ = N_- + \lfloor Q \rfloor +1$ positive ions as a function of $Q$, as obtained from the simulation results for $\Delta F$ (using eq.~\eqref{pQ}), and from the Coulomb gas theory.}
\label{DF}
\end{figure}

\subsection{Grand canonical averaging}
Our simulations aim to account for a channel connecting two reservoirs, therefore the number of ions in the system should be able to fluctuate, the chemical potential $\mu$ of the reservoirs being fixed. At equilibrium, we can account for these fluctuations by carrying out simulations at a fixed number of ions $(N_+,N_-)$ and then averaging the observables obtained for different $(N_+,N_-)$ with the grand canonical probabilities of having $(N_+,N_-)$ ions in the channel. We assume that we may extend this approach out of equilibrium and carry out the same type of averaging for the observables $N(Q)$ and $I_{\pm}$ even when the system is driven by an electric field. 

We must now determine how many different values of $(N_+,N_-)$ should be taken into account. In the absence of surface charge, the system tries to maintain electroneutrality, $N_+ = N_-$. Departing from electroneutrality by a unit charge has an energy cost which is approximately the dielectric self-energy of the unit charge: $E_s = \xi/(2x_T) = 19.6 ~k_BT$ with our simulation settings (see section 1.6). Thus, the probability of observing a non-neutral state is negligible and the only allowed fluctuations are those that keep the system neutral, that is fluctuations in the number of ion pairs. In the CB regime, the ion pairs behave essentially as ideal gas particles, thus the fluctuations in their number become gaussian in the thermodynamic limit, with variance $\Delta N^2 \sim N$. Hence, when there is no surface charge, the grand canonical averaging takes care of itself thanks to the equivalence of ensembles, and we may carry out simulations with a fixed number $N = 100$ of ion pairs. At $N = 100$ we are not well into the thermodynamic limit, but we do not expect small fluctuations in the number of ion pairs to significantly affect the observables of interest. 

Now in the presence of a surface charge $Q$, the system tries to get as close to electroneutrality as possible, meaning that $N_+ = N_- + \lfloor Q \rfloor$ or $N_+ = N_- + \lfloor Q \rfloor +1$. If $Q$ is close to a half-integer, these two values of $N_+$ may have very similar probabilities, while other values of $N_+$ are essentially forbidden, since breaking the quasi-electroneutrality has an energy cost which is again at least $E_s = 19.6 ~k_BT$. This means that the fluctuations in $N_+$ and $N_-$ are not gaussian, whatever the system size, so that there is no equivalence of ensembles in the thermodynamic limit. One should therefore carry out separate simulations for the two allowed values of $N_+$.

Let us denote $p(Q)$ the probability of the channel containing $N_+ = N_- + \lfloor Q \rfloor +1$ positive ions, as in section 2.3. It is given by
\begin{equation}
p(Q) = \frac{e^{-F( \lfloor Q \rfloor +1)}}{e^{-F( \lfloor Q \rfloor +1)}+e^{-F( \lfloor Q \rfloor) }} \equiv \frac{1}{1+e^{\Delta F(Q)}},
\label{pQ}
\end{equation}
where $F(q)$ is a shorthand for $F(N_-+q,N_-)$, the free energy at fixed particle numbers $(N_-+q,N_-)$. We determined the free energy difference $\Delta F$ from brownian dynamics simulations by a thermodynamic integration method~\cite{frenkel}. Consider a system containing $N_+$ positive ions, $N_-$ negative ions and an extra particle of charge $\lambda e$. Let $F(\lambda)$ be the free energy of such a system. The derivative of the free energy with respect to $\lambda$ can be computed as
\begin{equation}
\frac{\partial F (\lambda)}{\partial \lambda} = \frac{\int \d \{ x_i \} (\partial E(\{ x_i\},\lambda)/\partial \lambda)e^{-E(\{ x_i\},\lambda)}}{\int \d \{ x_i \} e^{-E(\{ x_i\},\lambda)}} = \left \langle \frac{\partial E(\{ x_i \}, \lambda)}{\partial \lambda} \right \rangle_{\lambda},
\end{equation}
where $\langle \cdot \rangle_{\lambda}$ denotes averaging over the dynamics with a given value of $\lambda$. 
The energy $E(\{ x_i\},\lambda)$ can be decomposed as
\begin{equation}
E(\{ x_i\},\lambda) = E (\{x_i\},0) + \lambda \cdot [ E (\{x_i\},1) - E (\{x_i\},0) ].
\end{equation}
Here we did not take into account the dielectric self-energy of the extra particle, which is proportional to $\lambda^2$, but it can be added \emph{a posteriori} in the free energy difference. Precisely, $\Delta F = E_s + F(\lambda = 1) - F(\lambda = 0)$, hence
\begin{equation}
\Delta F = E_s+\int_0^1 \d \lambda \frac{\partial F (\lambda)}{\partial \lambda} =E_s+ \int_0^1 \d \lambda \left \langle E (\{x_i\},1) - E (\{x_i\},0)\right \rangle_{\lambda}
\label{DeltaF}
\end{equation}

\begin{figure}
\centering
\includegraphics{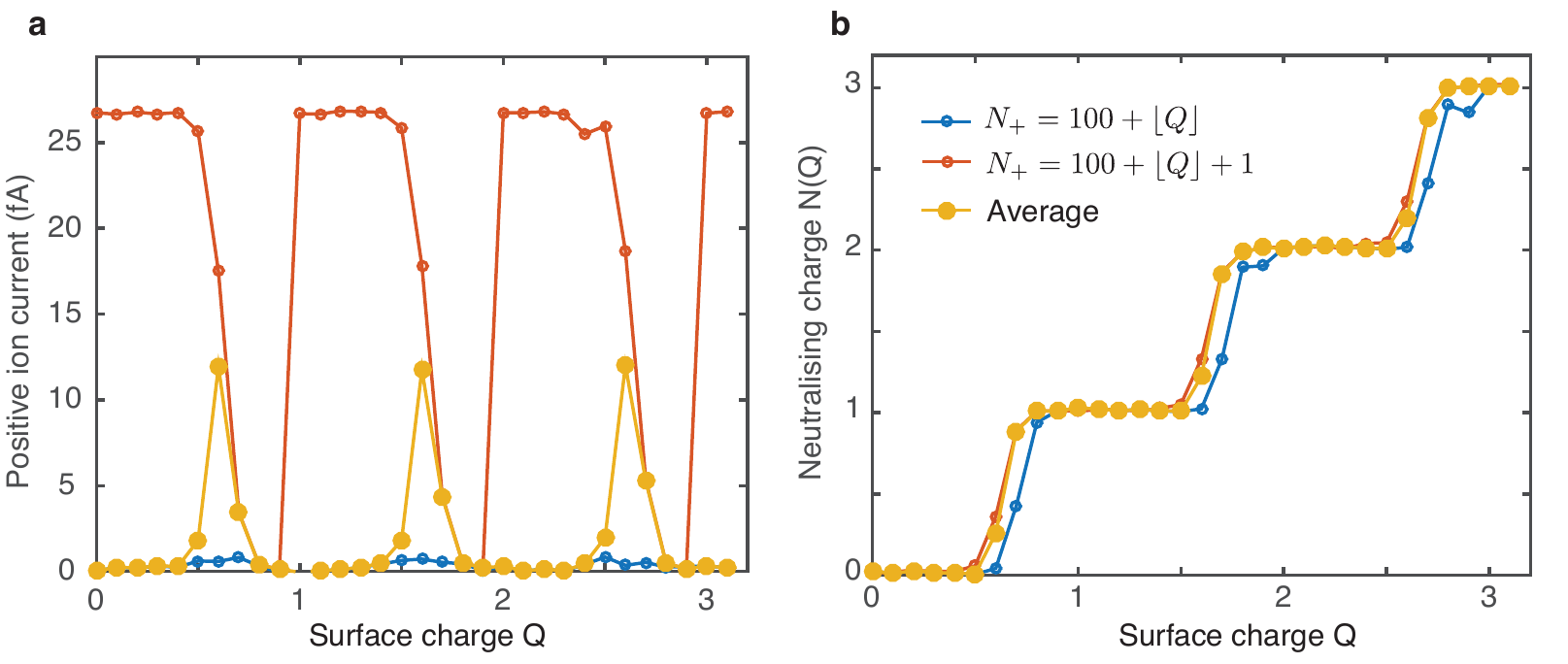}
\caption{\textbf{a.} Positive ion current as a function of surface charge, as obtained from brownian dynamics simulations, for each of the two allowed values of $N_+$, and the resulting grand-canonical average. \textbf{b.} Neutralising charge $N(Q)$ as a function of surface charge, as obtained from brownian dynamics simulations, for each of the two allowed values of $N_+$, and the resulting grand-canonical average.}
\label{avg}
\end{figure}

This expression suggests a numerical scheme for computing $\Delta F$. Simulations are carried out at a range of values of $\lambda$ between 0 and 1. For each simulation one computes the average of the quantity $E (\{x_i\},1) - E (\{x_i\},0)$. Concretely, this means that every few timesteps, we set $\lambda$ to 1, compute the energy, then set $\lambda$ to 0, compute the energy, take the difference between the two, then return $\lambda$ to its original value and continue the dynamics. We then numerically compute the integral in eq.~\eqref{DeltaF} to obtain $\Delta F$. We implemented this numerical scheme for a range of values of $Q$. For each value of $Q$ we sampled 20 values of $\lambda$, and each $5 \times 10^8$-timestep-long simulation was repeated 5 times, since the exploration of the configuration space is slower in the absence of driving by an electric field.

Figure~S\ref{DF} shows the results we obtained for $\Delta F$ as a function of $Q$, and the resulting probability $p(Q)$, with parameters $x_T = 0.9~\rm \AA$, $\xi = 3.5~\rm nm$, and $L = 1.25~\rm \mu m$ (a line density of 80 ions/$\mu$m). The simulation results for $p(Q)$ can be compared with the Coulomb gas theory. In the CB regime, any unpaired positive ions are most likely to be closely bound to the surface charge, hence it does not matter whether to consider the number of positive ions in the whole system or in the vicinity of the surface charge. Thus, as in section 2.3, $p(Q) = N(Q) - \lfloor Q \rfloor$, with $N(Q)$ given for example by eq.~\eqref{NQ}. Panel b shows that this analytical result is in excellent agreement with simulations. Thus, to carry out the grand canonical averaging in non-equilibrium simulations, we used the $p(Q)$ given by the Coulomb gas theory.

In order to give an idea of the effect of grand-canonical averaging, figure~S\ref{avg} shows the neutralising charge $N(Q)$ and the positive ion current $I^+$ obtained with each of the two allowed values of $N_+$, and the resulting average. 

\subsection{Ion pump}

\begin{figure}
\centering
\includegraphics{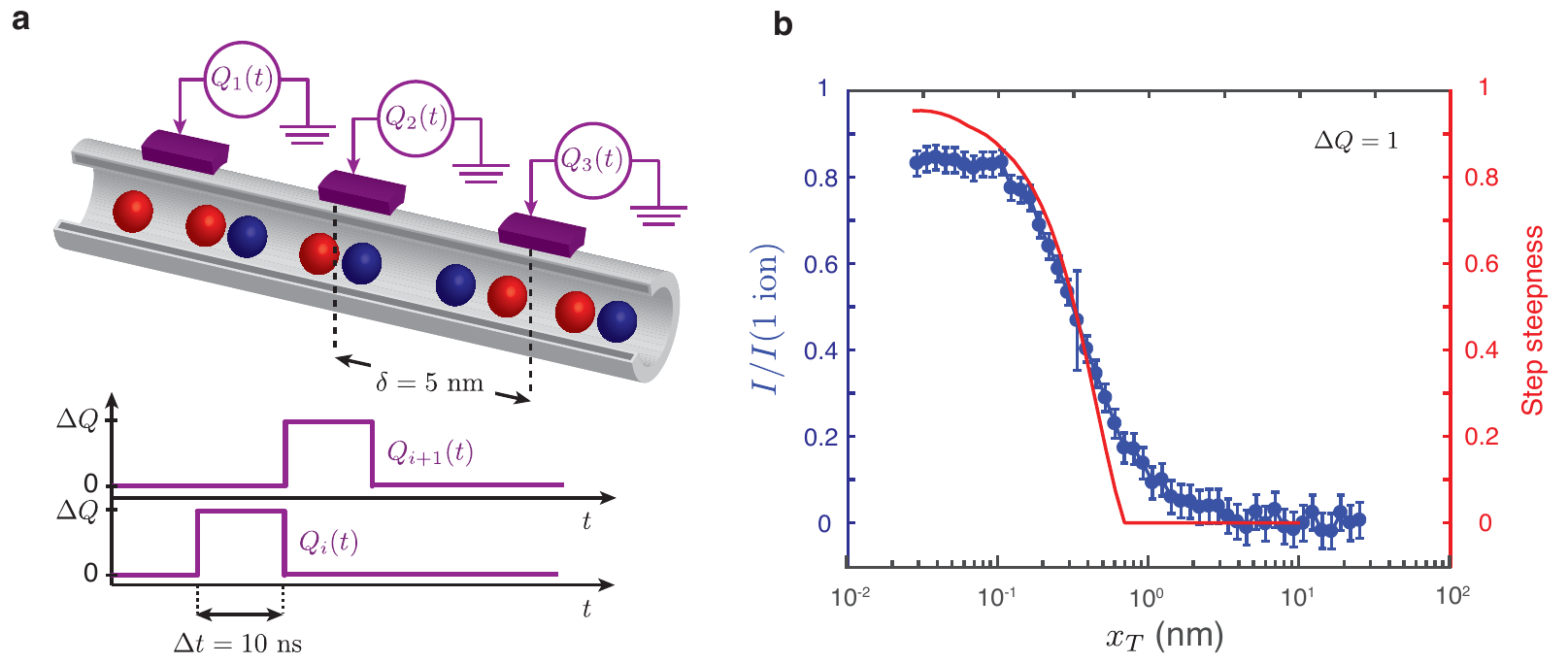}
\caption{\textbf{a.} Schematic of the ion pump and time dependence of the variable surface charges $Q_i(t)$. \textbf{b.} Positive ion current in the ion pump (blue circles), at fixed pumping amplitude $\Delta Q = 1$, as a function the thermal length $x_T$. The pumping current drops down to 0 for large values of $x_T$ (weak interactions), and correlates well with the CB step steepness obtained from the Coulomb gas theory (red line).}
\label{pump}
\end{figure}

We carried out a proof-of-concept simulation of a device that functions as an ion pump. The device consists of a nanochannel connected to not one, but several gating electrodes, which may induce time dependent surface charges on the channel walls. Its operating principle is shown in figure 4a of the main text, and recalled in figure S\ref{pump}. The variable charges are placed every 5 nm along the channel. At a given point in time, all the charges are set to 0, except one which takes the value $-\Delta Q$. Every 10 ns, the charge $-\Delta Q$ is moved to the next site, 5 nm further along the channel. We used a channel of length $L = 625~\rm nm$ with periodic boundary conditions. Intuitively, this displacement of a negative surface charge should result in the dragging of a positive ion along the channel. In figure 4b of the main text, we show the ionic current normalised by $I (\text{1 ion})$, the current that would result from the perfect pumping of a single ion, that is an ion moving at a velocity of 5 nm per 10 ns. It appears that the interactions need to be strong enough (that is $x_T$ low enough) for the pumping current to be close to the theoretical maximum. We carried out further simulations at fixed amplitude $\Delta Q = 1$, and a range of values of $x_T$, whose results are shown in figure S\ref{pump}b. We observe that the pumping current drops down to 0 at large values of $x_T$, and this decrease correlates well with step steepness obtained from the Coulomb gas theory at a given value of $x_T$. This highlights that the operation of the ion pump relies on the system being in the CB regime. 

For the simulation in figure S\ref{pump}b, with pumping amplitude $\Delta Q = 1$, we used a fixed number of particles: $N_- = 100$ and $N_+ = 101$. For the simulation in figure 4b of the main text, with $x_T = 0.1~\rm nm$, we used the same type of grand-canonical averaging as described in section 3.2. With $x_T = 2~\rm nm$, setting either $N_+ = N_- + \lfloor Q \rfloor$ or $N_+ = N_- + \lfloor Q \rfloor+1$ does not alter the result.

\newpage

\section{Movie legend}
\textbf{Movie S1: Fractional Wien effect as observed in simulations}. The movie shows 20 ns of a brownian dynamics simulation as described in section 3. The trajectories of positive ions (red spheres) and negative ions (blue spheres) are shown in the vicinity of the surface charge carried by the one-dimensional "channel", which is schematically depicted for clarity. The surface charge, set to $Q = -1.5$, initially binds two positive ions. Note that the other ions form tightly bound pairs. Starting at 0:06, a positive ion, pointed by a red arrow, unbinds from the surface charge and moves to the right under the effect of the electric field. When it encounters an ion pair, the free positive ion is exchanged through a Grotthus-like mechanism, so that another ion, still pointed by the red arrow, keeps moving the right, thus producing an electric current.
Parameters for the simulation are $x_T = 3.6~\AA$, $\xi = 3.5~\rm nm$, $E = 0.2~k_BT/\rm nm$ and a line density of 80 ions/$\mu$m. The ion trajectories are smoothed over 100 timesteps and visualised using VMD~\cite{vmd}.